\documentclass[preprint,12pt]{elsarticle}

\makeatletter
\def\ps@pprintTitle{%
 \let\@oddhead\@empty
 \let\@evenhead\@empty
 \def\@oddfoot{}
 \let\@evenfoot\@oddfoot}
\makeatother

\usepackage{geometry}
\usepackage{amsthm}
\usepackage{booktabs, rotating, multirow, array, multicol} %for cmidrule command in tables % for rotating text in a cell
\usepackage{caption, subcaption, rotating}
\usepackage{hyperref}
\hypersetup{
    colorlinks=true,
    linkcolor=cyan,
    filecolor=magenta,      
    urlcolor=cyan,
    pdftitle={Overleaf Example},
    pdfpagemode=FullScreen,
    }
\usepackage{amssymb}
\usepackage{amsmath}
\begin{document}

\begin{frontmatter}


\cortext[cor1]{Corresponding author}

\title{Observed Fisher Information in hidden Markov models \\ Application to a noisy Gaussian random walk}
 \author[label1]{Alexandra Lefebvre\corref{cor1}}
 \ead{alexandra.lefebvre@math.cnrs.fr}
 
 \author[label1]{Grégory Nuel} %% Author name

 \affiliation[label1]{organization={Sorbonne Université, CNRS},
                addressline={Laboratoire de Probabilités, Statistique et Modélisation (LPSM)}, 
                city={Paris},
               citysep={}, % Uncomment if no comma needed between city and postcode
                postcode={F-75005}, 
%                state={},
                country={France}}

%% Abstract
\begin{abstract}
In this work we provide analytical and closed-form expressions for the exact computation of the score and the observed Fisher information matrix in a Gaussian random walk observed through Gaussian noise. Our method is based on the Oakes' identity and, as for the computation of the log-likelihood, its complexity in time is linear in the length of the sequence with the forward-backward (or Baum-Welch) algorithm. We illustrate the method over various simulation studies and provide parameter estimates computed with the Newton-Raphson algorithm along with confidence intervals. 
\end{abstract}

%% Keywords

\end{frontmatter}

\section{Intro}

Hidden Markov models (HMMs) are a particular class of latent variable models introduced in the late 60's and early 70's \citep{baum1966statistical, baum1970maximization}. They received increasing attention for their particular dependence structure that allows for recursive computational methods based on the forward-backward (FB) algorithm \citep{baum1970maximization} and for their large range of applications including bioinformatics, signal processing, econometrics, time-series analysis, etc. Exact implementation is usually available in discrete HMM (discrete latent states and continuous or discrete observations) and Gaussian linear state-space models (Gaussian latent states and Gaussian observations) whereas numerical approximations are required for most other HMMs with continuous latent states. 

The derivatives of the log-likelihood are key elements in statistics, in particular the score, that is the first order derivatives of the log-likelihood and the observed Fisher information matrix (observed FIM) as the negative of the Hessian matrix (second order derivatives) of the log-likelihood. These quantities are useful in numerous statistical applications such as computing standard error of parameter estimates or performing statistical tests, to mention a few. It is also worth noticing that the Expectation-Maximization (EM) algorithm \citep{dempster1977maximum}, referred to as the Baum-Welch algorithm in the particular case of HMMs \citep{baum1970maximization}, do not require the derivatives of the log-likelihood. It is therefore a usual choice for likelihood maximization in latent variable models. Nevertheless alternative choices, such as the Newton-Raphson (NR) algorithm which can be faster and easier to implement in some cases, become available with the score and the observed FIM. 
 
The missing information principle of \citet{orchard1972missing}, that expresses the observed FIM as the difference between a matrix related to complete information and a matrix related to missing information du to the unobserved variables is the starting point of most methods for computing the observed FIM, notably Louis' identity \citep{louis1982finding} and Oakes' identity \citep{oakes1999direct}.  
Both identities, when applied to HMMs, display direct outputs of the classical FB algorithm (so-called smoothing and filtering), except one quantity of more complicated computation, that is, the sum of cross-products (respectively the gradient of posterior state densities) in Louis' identity (respectively Oakes' identity). The introduction of each identity was therefore followed by the development of computational methods for its numerical approximation or exact computation. 

Numerical approximations are inevitable in most state-space models, models of continuous states. They include Monte Carlo methods  
 \citep{turner1998hidden, delyon1999expectation} or finite difference approximation  \citep{chalmers2018numerical} to approach Louis' and Oakes’ identities respectively. 
Nevertheless, exact methods provide exact estimates, are more stable and computationally more efficient. They are therefore preferable whenever applicable, that is, in most discrete HMM and Gaussian linear state-space models. 
The formulas of \citet{cappe2005recursive} for instance can be recursively computed to obtain the sum of cross-product in Louis' identity with an additional computational cost limited to the square of the state space times the number of observations. Their exact implementation is relatively simple in a discrete HMM,
but seems, in our view, not trivial and not explicitly detailed by the authors for a Gaussian linear state-space model. 
The application of Oakes' identity seems, in our view, easier to implement and computationally more efficient as it simply requires first order derivatives of smoothing and filtering quantities. Moreover, it implies the gradient of posterior state distributions that can be useful for alternative interesting purpose. 
Whereas \citet{bartolucci2015information} presented an example of exact computation of Oakes' identity in a discrete HMM, the case of a Gaussian random walk observed through Gaussian noise involves non trivial closed-form expressions that we wish to present in this work. 

Among exact methods for computing the derivatives of the log-likelihood, it is also worth mentioning those based on the exact computation of the derivatives of smoothing functions detailed in \citep{lystig2002exact} for the second order derivatives and \citep{lefebvre2018sum} for the extension to higher orders. Nevertheless these are computationally more intensive than the aforementioned ones for the second order derivatives. Consequently their use seems only justified when higher order derivatives are required \citep{lefebvre2018sum}. 

This paper is organized as follows. The entire Section~\ref{sec:notation} is placed within the general HMM framework with no assumption on the finiteness of the state space. The rest of the work is related to the hidden Gaussian random walk with Gaussian noise model. In Section~\ref{sec:notation}, we firstly introduce the notation along with main forward and backward quantities involved in the classical FB algorithm. We secondly remind the expression of the score as in the Fisher's identity from which we deduce the observed FIM, as in Oakes' identity, applied to an HMM. We thirdly detail individual components of Oakes' identity as a combination of direct outputs of the classical FB algorithm. Whereas their implementation is straightforward for a discrete HMM, they involve non trivial analytical expressions within a Gaussian state space framework. We therefore provide in Section~\ref{sec:application} all analytical expressions and closed-forms of the quantities introduced in the previous section when applied to a noisy Gaussian random walk. In Section~\ref{sec:results} we propose an illustration of the method with the computation of the score and the observed FIM in different simulation studies. Finally we resume the method and discuss main perspectives in Section~\ref{sec:discussion}.

%%%%%%%%%%%%%%%%%%%%%
%NOTATION AND CONTEXT
%%%%%%%%%%%%%%%%%%%%%

\section{Notation and context}\label{sec:notation}

We consider a Hidden Markov Model (HMM) parametrized by $\theta$, composed of a hidden latent Markov chain $X = (X_1,\ldots, X_n)$ and observed variables $Z = (Z_1,\ldots, Z_n)$ such that the joint distribution of $X$ and $Z$ writes 
\begin{equation*}
\mathcal{P}(X,Z=z|\theta) = 
\rho(x_1|\theta) \prod_{i=2}^n \pi (x_{i-1}, x_i | \theta) \prod_{i=1}^n \varphi_i (x_i | \theta) 
\end{equation*}
where $z = (z_1,\ldots, z_n)$ is the observed vector $Z$, for $i \in \{1,\ldots, n\}$, $x_i$ is the value taken by $X_i$,
$$
\rho(x|\theta) = \mathcal P(X_1=x|\theta)
$$
$$
\text{for } i\in\{2,\ldots,n\},\quad
\pi(x, y | \theta) = \mathcal P(X_i = y | X_{i-1}=x; \theta)
$$
$$
\text{for } i\in\{1,\ldots,n\},\quad
\varphi_i(x | \theta) = \mathcal P(X_i=x | Z_i=z_i; \theta)
$$
with no assumption on probability distributions at this stage. Functions $\varphi_i$ are indexed with $i$ instead of conditioning on $\{Z_i=z_i\}$ for lightening the notation. They may therefore vary from one observed sequence to another one. Usually $\rho$, $\pi$ and $\{\varphi_i\}_{i>0}$ are respectively called initial, transition and emission densities. 

\subsection{The forward-backward algorithm}\label{sec:fb} 

In this paragraph, we briefly outline key parts of the FB algorithm of \citet{baum1970maximization} and we refer to the tutorial of \citet{rabiner1989tutorial} or, for instance, the book of \citet{cappe2005recursive} for more details. The forward and the backward quantities (FB quantities) are respectively defined, for $i \in \{1,\ldots, n\}$, as
$$
f_i(x|\theta) 
\stackrel{\text{def}}{=} 
\mathcal P(Z_1 = z_1, \ldots, Z_i=z_i, X_i=x | \theta)
$$
and 
$$
h_i(x|\theta) 
\stackrel{\text{def}}{=}
\mathcal P(Z_{i+1}=z_{i+1}, \ldots, Z_n=z_n | X_i=x; \theta).
$$ 
Following the FB algorithm, they
can be recursively computed with a process 
often referred to as smoothing and filtering in the HMM literature. 
Forward quantities are initialized with
\begin{equation}\label{eq:f1}
f_1(x|\theta) = \rho(x|\theta) \varphi_1 (x | \theta) 
\end{equation}
then recursively computed for $i = 2,\ldots,n$ with
\begin{equation}\label{eq:fi}
f_i(y|\theta) = 
\int f_{i-1}(x|\theta) \pi (x,y | \theta) \varphi_i (y | \theta)\,dx.
\end{equation}
Backward quantities are initialized, for all $x$ in the domain of $X_i$, with
\begin{equation}\label{eq:bn}
h_n(x) = 1 
\end{equation}
then for $i = n,\ldots,2$,
\begin{equation}\label{eq:bi}
h_{i-1}(x|\theta) = \int h_i(y|\theta)  \pi(x,y | \theta) \varphi_i (y | \theta) \,dy.
\end{equation}
As forward and backward quantities may reach values below machine precision with an increasing $n$, each quantity is usually normalized with an appropriate scaling factor to stay within a tractable bound. The most commonly used scaling factor is such that each FB quantity sums to one as suggested in the tutorial of \citet{rabiner1989tutorial}. 

Finally, the joint probability of $\{X_i, Z=z\}$ and of $\{X_{i-1}, X_i, Z=z\}$ is respectively given by
\begin{equation}
\mathcal{P}(X_i=x, Z=z|\theta) = f_i(x|\theta) h_i(x|\theta) 
\label{eq:joint.x}
\end{equation}
and 
\begin{equation}
\mathcal{P}(X_{i-1}=x, X_i=y, Z=z|\theta) = f_i(x|\theta) \pi (x,y | \theta) \varphi_i (y | \theta) h_i(y|\theta).
\label{eq:joint.xy}
\end{equation}
Therefore, one can obtain the likelihood $L(\theta) \stackrel{\text{def}}{=} \mathcal{P}(Z=z | \theta)$ by integrating $x$ out of Equation~(\ref{eq:joint.x}) or integrating $x$ and $y$ out of Equation~(\ref{eq:joint.xy}) and finally, the posterior distribution of latent states, by dividing Equation~(\ref{eq:joint.x}) or Equation~(\ref{eq:joint.xy}) by $L(\theta)$. Note that the forward pass is sufficient for the likelihood alone as, choosing the last index $n$ and Equation~(\ref{eq:joint.x}), we have in particular
\begin{equation}\label{eq:likelihood}
L(\theta) = \int f_n(x|\theta) \, dx.
\end{equation}
For the rest of the paper, we will denote the log-likelihood by $\log L(\theta) = \ell(\theta)$. 

\bigskip
As previously mentioned, the numerical evaluation of the integrals involved in the FB algorithm, in particular Equations~(\ref{eq:fi}) and~(\ref{eq:bi}), is exact with closed-form formulas in the particular cases of discrete HMM and, because the product of two Gaussian probability density functions is also Gaussian, in Gaussian-linear state space models (see for instance Chapter 5 of \citet{cappe2005inference} for details). However one usually need numerical approximation methods for other continuous state-space models.

%%%%%%%%%%%%%%%%%%%%%
% OBSERVED FIM
%%%%%%%%%%%%%%%%%%%%%

\subsection{Observed Fisher information matrix}

The observed Fisher information matrix is defined as the negative of the Hessian (or the matrix of second order derivatives) of the log-likelihood. 

Although the  EM algorithm \citep{dempster1977maximum} is classically used for parameter estimation in latent variable models, the authors and discussants leveraged additional interesting properties of its auxiliary function defined as 
$$
\mathcal Q(\theta|\theta') \stackrel{\text{def}}{=} \int \mathcal P(X=x | Z=z; \theta') \log \mathcal P(X=x, Z=z | \theta) \, dx.
$$
The auxiliary function of EM, in the framework of an HMM with above notation, becomes
\begin{multline}\label{eq:auxiliary.em}
\mathcal Q(\theta|\theta') = 
\int \mathcal P(X_1= x | Z=z; \theta') \log \rho(x|\theta) \ dx\\
+\sum_{i=1}^n \int \mathcal P(X_i=x|Z=z;\theta') \log \varphi_i(x|\theta) \ dx\\
+ \sum_{i=2}^n \int\int \mathcal P(X_{i-1}=x, X_i=y|Z=z;\theta') \log \pi(x, y|\theta) \, dxdy.
\end{multline}
A key identity within latent variable models, sometimes referred to as the Fisher's identity (see discussant in \citep{dempster1977maximum}), stands that, for any $\theta'\neq\theta$, the gradient of $\mathcal Q(\theta|\theta') $
with respect to $\theta$ and evaluated at $\theta'$ equals the score evaluated at $\theta'$. 
In other words, differentiating Equations~(\ref{eq:auxiliary.em}) with respect to $\theta$ and evaluating it at $\theta'$, the score writes
\begin{multline}\label{eq:score}
\nabla \ell(\theta) =  
\int \mathcal P(X_1= x | Z=z; \theta) \nabla \log \rho(x|\theta) \ dx\\
+\sum_{i=1}^n \int \mathcal P(X_i=x|Z=z;\theta) \nabla \log \varphi_i(x|\theta) \ dx\\
+ \sum_{i=2}^n \int\int \mathcal P(X_{i-1}=x, X_i=y|Z=z;\theta) \nabla \log \pi(x, y|\theta) \ dxdy
\end{multline}
after replacing $\theta'$ by $\theta$ in the final equation for readability. 
As a sketch of the proof, we have 
$
\mathcal Q(\theta|\theta') = \ell(\theta) - \mathcal H(\theta|\theta')
$
where $\mathcal H(\theta|\theta') = - \int \mathcal P(X|Z=z; \theta') \log \mathcal P(X=x|Z=z; \theta) \, dx$ and the gradient of $\mathcal H$ with respect to $\theta$ vanishes at $\theta'$ (see Chapter 10, in particular Proposition 10.1.6 of \citet{cappe2005inference} for more details). 

Finally, differentiating Equation~(\ref{eq:score}) once more with respect to $\theta$, the Hessian of the log-likelihood writes
\begin{align}\label{eq:hessian}\nonumber
\nabla^2 \ell(\theta) =  
\int \nabla \mathcal P(X_1&= x|Z=z; \theta) \{\nabla^t \log \rho(x|\theta)\} \ dx\\\nonumber
&+ \int \mathcal P(X_1= x | Z=z; \theta) \{\nabla^2 \log \rho(x|\theta)\} \ dx\\\nonumber
&+\sum_{i=1}^n \int \nabla \mathcal P(X_i=x|Z=z;\theta) \{\nabla^t \log \varphi_i(x|\theta)\} \ dx\\\nonumber
&+\sum_{i=1}^n \int \mathcal P(X_i=x|Z=z;\theta) \{\nabla^2 \log \varphi_i(x|\theta)\} \ dx\\\nonumber
&+ \sum_{i=2}^n \int\int \nabla \mathcal P(X_{i-1}=x, X_i=y|Z=z;\theta) \{\nabla^t \log \pi(x, y|\theta)\} \ dxdy\\
&+ \sum_{i=2}^n \int\int \mathcal P(X_{i-1}=x, X_i=y|Z=z;\theta) \{\nabla^2 \log \pi(x, y|\theta)\} \ dxdy
\end{align} 
where exponent $t$ stands for transpose and $\nabla^2$ for the Hessian. 

Note that the resulting Equation~(\ref{eq:hessian}) is equivalent to applying Oakes' identity \citep{oakes1999direct} according to which 
$
\nabla^2 \ell(\theta) = \left\{
\partial^2 \mathcal Q(\theta'|\theta) / \partial\theta'^2 + \partial^2 \mathcal Q(\theta'|\theta) / \partial\theta'\partial\theta
\right\}_{\theta'=\theta}
$
where $\{\cdot\}_{\theta'=\theta}$ stands for the value of $\{\cdot\}$ taken at $\theta'=\theta$. 

Provided that the first and second order derivatives of $\log \rho$, $\{\log\varphi_i\}_{i>0}$ and $\log \pi$ with respect to $\theta$ admit an analytical expression, the only quantities of Equation~(\ref{eq:hessian}) that were not introduced in Section~\ref{sec:fb} are $\nabla \mathcal{P}(X_i=x|Z=z; \theta)$ and $\nabla\mathcal{P}(X_{i-1}=x, X_{i}=y | Z=z; \theta)$. Nevertheless these quantities are 
easily expressed as functions of FB quantities along with their gradient and the score (the gradient of the log-likelihood). Indeed, for any subset $\bar X$ of $X = \{X_1,\ldots, X_n\}$, as $\mathcal P(\bar X | Z=z;\theta) = \mathcal P(\bar X, Z=z | \theta) / L(\theta)$, its gradient writes
\begin{equation}\label{eq:grad.post}
\nabla \mathcal P(\bar X |Z=z;\theta) 
= \frac{\nabla \mathcal P(\bar X,Z=z|\theta)}{L(\theta)} - \mathcal P(\bar X|Z=z;\theta) \nabla \log L(\theta).
\end{equation}
Equation~(\ref{eq:grad.post}) holds of course for $X_i$ and $\{X_{i-1}, X_i\}$ and it solely involves $\nabla \mathcal P(\bar X,Z=z|\theta)$ and $\nabla \log L(\theta)$ that were not previously detailed. 
Nevertheless, starting from Equations~(\ref{eq:joint.x}), (\ref{eq:joint.xy}) and~(\ref{eq:likelihood}), we can see that gradients $\nabla \mathcal P(X_i=x, Z=z|\theta)$, $\nabla \mathcal P(X_{i-1}=x,X_i=y, Z=z|\theta)$ and $\nabla \log L(\theta)$ can be expressed with FB quantities along with their gradients only.

Finally, returning to the recursive implementation of FB quantities in Equation~(\ref{eq:f1}), (\ref{eq:fi}), (\ref{eq:bn}) and~(\ref{eq:bi}), it is well-known and straightforward to see that a similar recursive implementation exists for computing their gradient with the same complexity in time, using the gradient of  initial, transition and emission densities $\nabla \rho$, $\nabla\pi$ and $\{\nabla\varphi_i\}_{i>0}$. 

\bigskip
Let us stress that, although the above method is rooted on the auxiliary function of the EM algorithm, it does not necessarily involves the EM algorithm for parameter estimation. Precisely one can compute the score and the observed FIM with this method to implement, for instance, the NR algorithm for parameter estimation. 

\subsection{Score} As mentioned just above, the gradient of the likelihood is simply given by integrating $x$ (or respectively $x$ and $y$) out of the gradient $\nabla \mathcal P(X_i=x, Z=z|\theta)$ (respectively $\nabla \mathcal P(X_{i-1}=x,X_i=y, Z=z|\theta)$) and both $\nabla \mathcal P(X_i=x, Z=z|\theta)$ and $\nabla \mathcal P(X_{i-1}=x,X_i=y, Z=z|\theta)$ can be expressed with the gradient of FB quantities using Equation~(\ref{eq:joint.x}) (respectively Equation~(\ref{eq:joint.xy})). In practice, as for the likelihood, only the forward recursion is needed for its gradient as one can choose Equation~(\ref{eq:joint.x}) and the last index $n$. Finally, dividing the result by the likelihood provides the score. 

\section{Application to a Gaussian random walk observed through Gaussian noise}\label{sec:application}

We deduce from the previous section that, provided that all functions $\rho$, $\pi$ and $\{\varphi_i\}_{i>0}$ admit an analytical expression for their first and second order derivatives with respect to the parameter, all quantities involved in Equation~(\ref{eq:hessian}) and in the differentiation of Equation~(\ref{eq:likelihood}) admit a closed-form expression in a discrete HMM (discrete state space and no assumption on the finiteness of the observations). 
Note that considering continuous observations brings no more difficulty as integrations take place over latent variables. 

The case of a latent Gaussian random walk observed through Gaussian noise is slightly more complicated and the purpose of this section is to provide closed-form expressions of all quantities needed to adapt the above method to such models. 

\subsection{Notation and forward-backward recursion (smoothing, filtering).}
Let us consider the univariate Gaussian random walk with Gaussian observation noise composed of a latent Markov chain $X = (X_1,\ldots,X_n)$ and observed variables $Z = (Z_1,\ldots,Z_n)$ such that $X_1$ is univariate Gaussian of mean $\alpha$ and variance $\varepsilon^2$ , for $i\in\{2,\ldots,n\}$
$$
X_i = X_{i-1} + \varepsilon U_{i-1}
$$
and for $i\in\{1,\ldots,n\}$
$$
Z_i = X_i + \eta V_{i}
$$
where $\{U_i\}_{i\in\{1,\ldots,n-1\}}$ and $\{V_i\}_{i\in\{1,\ldots,n\}}$ are independent and identically distributed standard Gaussian variables. This model is a particular case of an HMM and therefore, all previous equations are valid with 
\begin{equation}\label{eq:local.densities}
\rho(x|\theta) = g(x,\alpha, \varepsilon),
\quad
\pi(x,y|\theta) = g(y,x,\varepsilon)
\quad \text{and} \quad
\varphi_i(x|\theta) = g(z_i,x,\eta)
\end{equation}
where $g(., \mu, \sigma)$ denotes the Gaussian density of mean $\mu$ and variance $\sigma^2$. 

As noted in the literature, it is common to assume that $\rho$ is either fixed or fully determined by $\eta$ and $\varepsilon$ (see for instance Section 10.2.2 and 10.3.4 in \citep{cappe2005inference}). 
We assume here that $\alpha$ is fixed, therefore $\rho$ is parametrized by $\varepsilon$. The model is parametrized by $\theta = (\theta_1, \theta_2) = (\log \eta, \log \varepsilon)$.  Considering the logarithm of $\eta$ and $\varepsilon$ instead of $\eta$ and $\varepsilon$ leads to simpler analytical expressions of the derivatives of the log-likelihood. 

As previously mentioned, because the product of two Gaussian densities is also a Gaussian density, the particular case of both $X$ and $Z$ Gaussian implies Gaussian forward and backward quantities. 
Indeed, based on the following identity 
\begin{equation}\label{eq:prod}
g(x,\mu_1,\sigma_1)g(x,\mu_2,\sigma_2) = 
g\left(\mu_1,\mu_2,\sqrt{\sigma_1^2+\sigma_2^2}\right)
g\left(
x, \frac{\mu_2 \sigma_1^2 + \mu_1 \sigma_2^2}{\sigma_1^2+\sigma_2^2},\sqrt{\frac{\sigma_1^2\sigma_2^2}{\sigma_1^2+\sigma_2^2}}
\right)
\end{equation}
developed and proven in details in the internal report of \citet{bromiley2003products}, we can show that forward quantities can be written and recursively computed in the form of 
\begin{equation}\label{eq:fw}
f_i(x) = e^{K_i} g(x,\mu_i,\sigma_i)
\end{equation}
with 
$$
K_1 = \log g(z_1,\alpha,\sqrt{\varepsilon^2+\eta^2}), \quad
\mu_1 = \frac{z_1\varepsilon^2+\alpha\eta^2}{\varepsilon^2+\eta^2}, \quad
\sigma_1 = \sqrt{\frac{\varepsilon^2\eta^2}{\varepsilon^2+\eta^2}}
$$
and, for $i=2,\ldots,n$, 
$$
K_i = K_{i-1} + \log g(z_i,\mu_{i-1},\sqrt{\sigma_i^2+\eta^2}),
$$
where
$$
\mu_i = \frac{z_i \gamma_i^2 + \mu_{i-1} \eta^2}{ \gamma_i^2+ \eta^2}, \quad
\sigma_i=\sqrt{\frac{\gamma_i^2\eta^2}{\gamma_i^2+\eta^2}}
\quad\text{and}\quad
\gamma_i = \sqrt{\sigma_{i-1}^2+\varepsilon^2}.
$$
Similarly backward quantities can be written and recursively computed in the form of
\begin{equation}\label{eq:bk}
h_i(x) = e^{L_i}g(x,\nu_i,\tau_i)
\end{equation}
where
$$
L _n = -\log g(0,\nu_n,\tau_n) \quad\text{with}\quad
\nu_n=0, \quad \tau_n=+\infty
$$
and for $i=n,\ldots,2$, 
$$
L_{i-1} = L_i+\log \, g\left(z_i,\nu_i,\sqrt{\eta^2+\tau_i^2}\right)
$$
with
$$
\nu_{i-1}=\frac{\nu_i\eta^2+z_i\tau_i^2}{\eta^2+\tau_i^2}, \quad
\tau_{i-1}=\sqrt{\varepsilon^2 + \omega_i^2 }
\quad\text{and}\quad
\omega_i = \sqrt{\frac{\eta^2\tau_i^2}{\eta^2+\tau_i^2}}
$$ 
(see proof of Equations~(\ref{eq:fw}) and~(\ref{eq:bk}) along with recursive formulas in~\ref{app:proof.eq.fw.bk}). Although all quantities $\{f_i\}_{i>0}$, $\{h_i\}_{i>0}$, $\{K_i\}_{i>0}$, $\{L_i\}_{i>0}$, $\{\mu_i\}_{i>0}$, $\{\sigma_i\}_{i>0}$, $\{\nu_i\}_{i>0}$ and $\{\tau_i\}_{i>0}$ are parametrized by $\theta$, the parameter is dropped from the notation in the following for a better readability. Note that $K_i$ and $L_i$ play the role of scaling factor mentioned in Section~\ref{sec:fb} to keep FB quantities in a tractable bound and avoid computational underflow. 

Setting initial standard deviation at $\tau_n=10^5$ for the practical implementation of $L_n$ is large enough in most cases for approximating $B_n(y)\simeq 1$ for all $y \in \mathbb R$. 

Based on Equation~(\ref{eq:joint.x}) and Equation~(\ref{eq:joint.xy}) respectively, it is finally clear that the posterior distribution of 
$X_i$ writes
\begin{equation}\label{eq:post.x}
\mathcal{P}(X_i=x|Z=z; \theta) = 
g(x, M_i, S_i)
\end{equation}
with
$$
M_i = \frac{\nu_i\sigma_i^2+\mu_i\tau_i^2}{\sigma_i^2+\tau_i^2}
\quad \text{and} \quad 
S_i = \sqrt{\frac{\sigma_i^2\tau_i^2}{\sigma_i^2+\tau_i^2}}
$$
and the posterior distribution of $\{X_{i-1},X_i\}$ for all $i>1$ is given by
\begin{equation} \label{eq:post.xy}
\mathcal{P}(X_{i-1} = x, X_i=y | Z=z; \theta)
= g(x, P_i y+Q_i, R_i)\ g(y,N_i, T_i)
\end{equation}
with
$$
P_i = \frac{\sigma_{i-1}^2}{\gamma_i^2}, \quad
Q_i = \mu_{i-1} \frac{\varepsilon^2}{\gamma_i^2}, \quad
R_i = \sqrt{P_i\varepsilon^2}
$$
and 
$$
N_i= \frac{
\nu_{i-1}\gamma_i^2 + \mu_{i-1} \omega_i^2
}{
\omega_i^2 + \gamma_i^2
}, \quad 
T_i = \sqrt{ \frac {
\omega_i^2 \gamma_i^2
}{
\omega_i^2 + \gamma_i^2
}}
$$
where $\gamma_i$ and $\omega_i$ are introduced in Equations~(\ref{eq:fw}) and~(\Ref{eq:bk}) respectively (see proof of Equations~(\ref{eq:post.x}) and~(\ref{eq:post.xy}) in~\ref{app:proof.eq.post.xy}).

\subsection{Gradient of forward and backward quantities and gradient of posterior state densities}

We remind that the Hessian of the log-likelihood as expressed in Equation~(\ref{eq:hessian}) displays the gradient of posterior state densities for which we provide analytical expressions in this section. For that purpose, let us start with the gradient of FB quantities. 

We can show that the gradient of FB quantities with respect to $\theta = (\theta_1, \theta_2) = (\log \eta, \log \varepsilon)$ can be written and recursively computed in the form of
\begin{equation}\label{eq:grad.fb}
\nabla f_i(x) = 
A_i (x^k)_{k=0}^2 \ f_i(x)
\quad\text{and}\quad
\nabla h_i(x) = 
B_i (x^k)_{k=0}^2 \ h_i(x)
\end{equation}
where $(x^k)_{k=0}^2 = \begin{pmatrix} 1 & x & x^2\end{pmatrix}^t$ and $A_i$ and $B_i$ are $2\times 3$ matrices. The initial forward matrix $A_1$ writes
$$
A_1 = 
\begin{pmatrix}
-1 + z_1^2e^{-2\theta_1} & - 2 z_1 e^{-2\theta_1} & e^{-2\theta_1} \\
-1 + \alpha^2e^{-2\theta_2} & - 2 \alpha e^{-2\theta_2} & e^{-2\theta_2}
\end{pmatrix}
$$
and for $i=2,\ldots,n$,
\begin{multline*}
A_i = 
A_{i-1}
\begin{pmatrix}
1&0&0 \\
Q_i&P_i&0\\
Q_i^2+R_i^2& 2P_iQ_i& P_i^2
\end{pmatrix}\\
+
\begin{pmatrix}
-1 + z_i^2e^{-2\theta_1} & -2z_ie^{-2\theta_1} & e^{-2\theta_1}\\
-1 + (Q_i^2+R_i^2)e^{-2\theta_2} & 2(P_i-1)Q_ie^{-2\theta_2} & (P_i-1)^2e^{-2\theta_2}
\end{pmatrix}.
\end{multline*}
The initial backward matrix $B_n$ is a zero matrix of size $2\times3$ and for $i=n,\ldots,2$,
\begin{multline*}
B_{i-1} = 
B_i
\begin{pmatrix}
1&0&0 \\
\widetilde Q_i&\widetilde P_i&0\\
\widetilde Q_i^2+\widetilde R_i^2& 2\widetilde P_i\widetilde Q_i& \widetilde P_i^2
\end{pmatrix}\\
+
\begin{pmatrix}
-1 + \left[(\widetilde Q_i-z_i)^2+\widetilde R_i^2\right]e^{-2\theta_1} & 2\widetilde P_i(\widetilde Q_i-z_i)e^{-2\theta_1} & \widetilde P_i^2e^{-2\theta_1}\\
-1 + (\widetilde Q_i^2+\widetilde R_i^2)e^{-2\theta_2} & 2(\widetilde P_i-1)\widetilde Q_ie^{-2\theta_2} & (\widetilde P_i-1)^2e^{-2\theta_2}
\end{pmatrix}
\end{multline*}
with 
$$
\widetilde P_i = \frac{\omega_i^2}{\varepsilon^2 + \omega_i^2},
\quad
\widetilde Q_i = \frac{\nu_{i-1}\varepsilon^2}{\varepsilon^2 + \omega_i^2},
\quad
\widetilde R_i=\sqrt{\widetilde P_i  \varepsilon^2}
$$
(see proof of Equation~(\ref{eq:grad.fb}) along with the recursive formulas of matrices $\{A_i\}_{i\in\{1,\ldots,n\}}$ and $\{B_i\}_{i\in\{n,\ldots,1\}}$ in~\ref{app:proof.eq.grad}). 

We therefore deduce from Equation~(\ref{eq:grad.post}) and~(\ref{eq:grad.fb}) that the gradient of posterior state probabilities writes 
\begin{equation}\label{eq:grad.post.x}
\nabla \mathcal{P}(X_i=x | Z=z; \theta) = 
\left\{\left(A_i+B_i\right) (x^k)_{k=0}^2 - \nabla \ell(\theta) \right\}
g(x,M_i,S_i)
\end{equation}
and for $i>1$,
\begin{multline}\label{eq:grad.post.xy}
\nabla \mathcal{P}(X_{i-1}=x, X_{i}=y | Z=z; \theta) = \\
\left\{
A_{i-1} (x^k)_{k=0}^2 
 +
 \begin{bmatrix}
- 1+(y-z_i)^2 e^{-2\theta_1} \\
- 1+(y-x)^2 e^{-2\theta_2}
\end{bmatrix} 
+ B_i (y^k)_{k=0}^2  
- \nabla \ell(\theta) \right\} \\
\times g(x, P_i y+Q_i, R_i) g(y,N_i, T_i)
\end{multline}
(see proof of Equations~(\ref{eq:grad.post.x}) and~(\ref{eq:grad.post.xy}) in~\ref{app:proof.grad.post}).

\subsection{Score}

As mentioned in Section~\ref{sec:notation}, one can simply integrate the gradient of the last forward quantity over $x$ and divide the result by the likelihood to obtain the score. Therefore, in our context, the score writes 
\begin{equation}\label{eq:grad.ll}
\nabla \ell(\theta)  = A_n  
\begin{pmatrix}
1,& \mu_n,& \mu_n^2 + \sigma_n^2
\end{pmatrix}^t
\end{equation}
\begin{proof}
As $\nabla \ell(\theta) = \left\{\nabla \int F_n(x) \,dx\right\} / \left\{\int F_n(x) \,dx\right\}$, we have 
$$
\nabla \ell(\theta) = 
\frac{e^{K_n}}{e^{K_n}}
A_n
\int
(x^k)_{k=0}^2
\ g(x,\mu_n,\sigma_n) \,dx 
$$
and we extract the zero, first and second order moments of the Gaussian density of mean $\mu_n$ and variance $\sigma_n^2$.
\end{proof}

\subsection{Observed Fisher observation matrix}

In this section, we report individual components of the Hessian matrix of the log-likelihood for lightening the reading and we remind that the observed FIM is the negative of the Hessian. 
For any matrix $M$, we denote by $(M)_{j\cdot}$ the $j^\text{th}$ row of matrix $M$. 
\paragraph{First diagonal component of the observed FIM: $\boldsymbol{ -\partial^2 \ell(\theta) / \partial \theta_1^2}$} 

The first component of the Hessian matrix of the log-likelihood as detailed in Equation~(\Ref{eq:hessian}) writes
\begin{multline*}
\frac{\partial^2\ell(\theta)}{\partial \theta_1^2} = 
\sum_{i=1}^n 
\int 
\frac{\partial}{\partial \theta_1} \mathcal{P}(X_i=x | Z=z;\theta) 
\frac{\partial}{\partial \theta_1} \log g(x, z_i, \eta)  
\ dx
\\
+ 
\sum_{i=1}^n 
\int 
\mathcal{P}(X_i=x | Z=z;\theta) \frac{\partial^2}{\partial \theta_1^2} \log g(x, z_i, \eta)
\ dx.
\end{multline*}
Replacing the first and second order derivatives of $\log g(x, z_i, \eta)$ with respect to $\theta_1 = \log \eta$ by their analytical expression, the posterior density of $X_i$ by its expression in Equation~(\ref{eq:post.x}) and its partial derivative with respect to $\theta_1$ by its related component in Equation~(\Ref{eq:grad.post.x}), we can write
\begin{multline*}
\frac{\partial^2\ell(\theta)}{\partial \theta_1^2}
= 
\sum_{i=1}^n \int 
\left(A_i+B_i\right)_{1\cdot} \left(x^k\right)_{k=0}^2
\left(-1+(x-z_i)^2 e^{-2\theta_1} \right) 
 g(x,M_i,S_i)
\ dx
\\
-\sum_{i=1}^n \int  2(x-z_i)^2 e^{-2\theta_1}  g(x,M_i,S_i) \ dx
\end{multline*}
where $\left(A_i+B_i\right)_{1\cdot}$ denotes the first row of the matrix $\left(A_i+B_i\right)$.
Combining coefficients according to the powers of $x$, we get 
$$
\frac{\partial^2\ell(\theta)}{\partial \theta_1^2} = 
\sum_{i=1}^n \int 
C_i^t (x^k)_{k=0}^4
\ g(x,M_i,S_i) \,dx 
$$
where the vector $C_i$ is given by 
\begin{multline*}
C_i = 
\begin{pmatrix}
\{a_i+b_i\}_{10} - \frac{\partial\ell(\theta)}{\partial \theta_1} & 0&0\\
\{a_i+b_i\}_{11} & \{a_i+b_i\}_{10}  - \frac{\partial\ell(\theta)}{\partial \theta_1} & 0\\
\{a_i+b_i\}_{12} & \{a_i+b_i\}_{11} & \{a_i+b_i\}_{10}  - \frac{\partial\ell(\theta)}{\partial \theta_1}\\
0 & \{a_i+b_i\}_{12} & \{a_i+b_i\}_{11} \\
0 & 0 & \{a_i+b_i\}_{12}
\end{pmatrix}
\begin{pmatrix}
-1 + z_i^2e^{-2\theta_1} \\
-2z_ie^{-2\theta_1} \\
e^{-2\theta_1}
\end{pmatrix}\\
- 2 e^{-2\theta_1}
\begin{pmatrix}
z_i^2 \\ -2z_i \\ 1 \\ 0 \\ 0
\end{pmatrix}
\end{multline*}
and $\{a_i+b_i\}_{k\ell}$ denotes the component of the $k^\text{th}$ row and $\ell^\text{th}$ column of the matrix $(A_i+B_i)$.
We finally extract the zero to the fourth moments of the Gaussian distribution of mean $M_i$ and standard deviation $S_i$ to get the first component of the Hessian matrix 
\begin{equation}\label{eq:h11}
\frac{\partial^2\ell(\theta)}{\partial \theta_1^2} 
= \sum_{i=1}^n C_i^t 
\begin{pmatrix}
1, & 
M_i,&
M_i^2+S_i^2,&
M_i^3 + 3M_iS_i^2,&
M_i^4 + 6 M_i^2 S_i^2 + 3 S_i^4
\end{pmatrix}^t.
\end{equation}

\paragraph{Off diagonal component of the observed FIM: $\boldsymbol{ -\partial^2 \ell(\theta) / \partial \theta_2\partial \theta_1}$} We propose to explore the below diagonal component of the Hessian, that is differentiating the log-likelihood with respect to $\theta_1$ first, then $\theta_2$ as it leads to simpler equations. Of course proceeding first with respect to $\theta_2$ then $\theta_1$ leads to same exact results. 

The extraction of $\partial^2 \ell(\theta) / \partial \theta_2\partial \theta_1$ from the Hessian expressed in Equation~(\ref{eq:hessian}) gives
$$
\frac{\partial^2\ell(\theta)}{\partial \theta_2\partial \theta_1} 
= 
\sum_{i=1}^n \int 
\frac{\partial}{\partial \theta_2} \mathcal P(X_i=x|Z=z; \theta)
\frac{\partial}{\partial \theta_1} \log g(x,z_i,\eta) \ dx
$$
With a similar reasoning as the one in the previous paragraph we can write 
\begin{align}\label{eq:h21}
\frac{\partial^2\ell(\theta)}{\partial \theta_2\partial \theta_1}  
\nonumber
& = 
\sum_{i=1}^n \int 
\left(
(A_i+B_i)_{2\cdot} 
\left(x^k\right)_{k=0}^2 - \frac{\partial\ell(\theta)}{\partial \theta_2} \right) 
\left(-1+(x-z_i)^2 e^{-2\theta_1}\right)
g(x,M_i,S_i) \ dx\\
\nonumber
& = 
\sum_{i=1}^n \int 
D_i^t (x^k)_{k=0}^4
\ g(x,M_i,S_i) \,dx \\
\frac{\partial^2\ell(\theta)}{\partial \theta_2\partial \theta_1}  
& = 
\sum_{i=1}^n D_i^t 
\begin{pmatrix}
1, & 
M_i,&
M_i^2+S_i^2,&
M_i^3 + 3M_iS_i^2,&
M_i^4 + 6S_i^2 M_i^2 + 3 S_i^4
\end{pmatrix}^t
\end{align}
where the vector $D_i$ writes
$$
D_i^t = 
\begin{pmatrix}
\{a_i+b_i\}_{20} - \frac{\partial\ell(\theta)}{\partial \theta_2} & 0&0\\
\{a_i+b_i\}_{21} & \{a_i+b_i\}_{20}  - \frac{\partial\ell(\theta)}{\partial \theta_2} & 0\\
\{a_i+b_i\}_{22} & \{a_i+b_i\}_{21} & \{a_i+b_i\}_{20}  - \frac{\partial\ell(\theta)}{\partial \theta_2}\\
0 & \{a_i+b_i\}_{22} & \{a_i+b_i\}_{21} \\
0 & 0 & \{a_i+b_i\}_{22}
\end{pmatrix}
\begin{pmatrix}
-1 + z_i^2e^{-2\theta_1} \\
-2z_i e^{-2\theta_1} \\
e^{-2\theta_1}
\end{pmatrix}.
$$

\paragraph{Second diagonal component of the observed FIM: $\boldsymbol{ -\partial^2 \ell(\theta) / \partial \theta_2^2}$}

The extraction of the second diagonal component of the Hessian matrix of Equation~(\ref{eq:hessian}) is
\begin{multline}\label{eq:general.h22}
\frac{\partial^2\ell(\theta)}{\partial \theta_2^2}= 
\int 
\frac{\partial}{\partial \theta_2} \mathcal{P}(X_1=x|Z=z; \theta) \frac{\partial}{\partial \theta_2} \log g(x, \alpha, \varepsilon) \ dx 
\\
\hspace{-5.5em}
+ \int
\mathcal{P} (X_1=x|Z=z; \theta) \frac{\partial^2}{\partial \theta^2_2} \log g(x, \alpha, \varepsilon) \ dx 
\\
\hspace{8em}
+ \sum_{i=2}^n \int \int 
\frac{\partial}{\partial \theta_2} \mathcal{P}(X_{i-1}=x, X_{i}=y | Z=z; \theta)
\frac{\partial}{\partial \theta_2} \log g(y, x, \varepsilon) \ dxdy 
\\ 
+ \sum_{i=2}^n \int \int 
\mathcal{P}(X_{i-1}=x, X_{i}=y|Z=z; \theta)
\frac{\partial^2}{\partial \theta^2_2} \log g(y, x, \varepsilon)
 \ dxdy 
\end{multline}
For clarity, we will use notation $\frac{\partial^2\ell(\theta)}{\partial \theta_2^2}  = \left\{\frac{\partial^2\ell(\theta)}{\partial \theta_2^2} \right\}_1 + \left\{\frac{\partial^2\ell(\theta)}{\partial \theta_2^2}\right\}_{2\ldots n}$ where $\left\{\frac{\partial^2\ell(\theta)}{\partial \theta_2^2}\right\}_{1}$ and $\left\{\frac{\partial^2\ell(\theta)}{\partial \theta_2^2}\right\}_{2\ldots n}$ denote respectively the first two lines and the last two lines of Equation~(\ref{eq:general.h22}). Reproducing the same reasoning as for the previous two paragraphs, the first two lines of Equation~(\ref{eq:general.h22}) become
\begin{multline*} 
\left\{\frac{\partial^2\ell(\theta)}{\partial \theta_2^2}\right\}_1 =
\int 
\left(
\left(A_1+B_1\right)_{2\cdot} 
\left(x^k\right)_{k=0}^2 
- \frac{\partial\ell(\theta)}{\partial \theta_2}
\right)
\left(-1+(x-\alpha)^2 e^{-2\theta_2} \right) 
g(x,M_1,S_1) \ dx
\\
- \int 
2(x-\alpha)^2 e^{-2\theta_2}
g(x,M_1,S_1) \ dx
\end{multline*}
where $\left(A_1+B_1\right)_{2\cdot}$ denotes the second row of the matrix $\left(A_1+B_1\right)$.
Combining coefficients according to the powers of $x$, then extracting the first moments of the Gaussian density $g(\cdot,M_1, S_1)$ we can write 
\begin{align}\nonumber\label{eq:h22.1}
\left\{\frac{\partial^2\ell(\theta)}{\partial \theta_2^2}\right\}_1 
& =
\int 
E_1^t \left(x^k\right)_{k=0}^4 \ g(x,M_1,S_1) 
\ dx \\
& = 
E_1^t
\begin{pmatrix}
1, & 
M_1,&
M_1^2+S_1^2,&
M_1^3 + 3M_1S_1^2,&
M_1^4 + 6S_1^2 M_1^2 + 3 S_1^4
\end{pmatrix}^t
\end{align}
where 
\begin{multline*}
E_1 = 
\begin{pmatrix}
\{a_1+b_1\}_{20} - \frac{\partial\ell(\theta)}{\partial \theta_2} & 0&0\\
\{a_1+b_1\}_{21} & \{a_1+b_1\}_{20} - \frac{\partial\ell(\theta)}{\partial \theta_2} & 0\\
\{a_1+b_1\}_{22} & \{a_1+b_1\}_{21} & \{a_1+b_1\}_{20} - \frac{\partial\ell(\theta)}{\partial \theta_2}\\
0 & \{a_1+b_1\}_{22} & \{a_1+b_1\}_{21} \\
0 & 0 & \{a_1+b_1\}_{22}
\end{pmatrix}
\begin{pmatrix}
-1 + \alpha^2e^{-2\theta_2} \\
-2\alpha e^{-2\theta_2} \\
e^{-2\theta_2}
\end{pmatrix}\\
- 2 e^{-2\theta_2}
\begin{pmatrix}
\alpha^2 \\ -2\alpha \\ 1 \\ 0 \\ 0
\end{pmatrix}.
\end{multline*}

Finally the last two lines of Equation~(\ref{eq:general.h22}) write 
\begin{multline*}
\left\{\frac{\partial^2\ell(\theta)}{\partial \theta_2^2}\right\}_{2,\ldots,n} = 
\sum_{i=2}^n \int \int  
\frac{\partial}{\partial \theta_2} \mathcal{P}(X_{i-1}=x, X_{i}=y | Z=z; \theta)
\frac{\partial}{\partial \theta_2} \log g(y, x, \varepsilon)
\ dxdy
\\
+ 
\sum_{i=2}^n \int \int 
\mathcal{P}(X_{i-1}=x, X_{i}=y|Z=z; \theta)
\frac{\partial^2}{\partial \theta^2_2} \log g(y, x, \varepsilon)) 
\ dxdy.
\end{multline*}
Replacing the partial derivative of the posterior density of $\{X_{i-1}, X_i\}$ by its related analytical expression in Equation~(\ref{eq:grad.post.xy}) and the first and second order derivatives of $\log g(y, x, \varepsilon)$ with respect to $\theta_2$ by their respective analytical expression, we can write
\begin{multline*}
\left\{\frac{\partial^2\ell(\theta)}{\partial \theta_2^2}\right\}_{2,\ldots,n} = 
\sum_{i=2}^n \int \int 
\left(
\left(A_{i-1}\right)_{2\cdot} \left(x^k\right)_{k=0}^2 
- 1+ (y-x)^2 e^{-2\theta_2} 
+ \left(B_i\right)_{2\cdot} \left(y^k\right)_{k=0}^2  
- \frac{\partial \ell(\theta)}{\partial \theta_2}
\right) 
\\
\hspace{08em}
\times
\left(-1 + (y-x)^2e^{-2\theta_2} \right) 
\ g(x, P_i y+Q_i, R_i) g(y,N_i, T_i)
\ dxdy\\
-\sum_{i=2}^n \int \int  
2 (y-x)^2e^{-2\theta_2} 
\ g(x, P_i y+Q_i, R_i) g(y,N_i, T_i)
\ dxdy
\end{multline*}
After combining coefficients according to the powers of $x$ and $y$ we get
\begin{equation}\label{eq:h22.2}
\left\{\frac{\partial^2\ell(\theta)}{\partial \theta_2^2}\right\}_{2,\ldots,n} 
= 
\sum_{i=2}^n
\sum_{\substack{j,k \\ j+k\leq4}}
e_{i,jk} \int \int  \,
x^jy^k g(x,P_i y + Q_i , R_i) g(y,N_i,T_i) \,dxdy
\end{equation}
where components $(e_{i,jk})_{j+k\leq4}$ are given by 
$$
\begin{pmatrix}
e_{i,00}\\
e_{i,10}\\
e_{i,01}\\
e_{i,20}\\
e_{i,02}\\
e_{i,11}\\
e_{i, 30}\\
e_{i, 03}\\
e_{i, 21}\\
e_{i, 12}\\
e_{i,40}\\
e_{i,04}\\
e_{i, 31}\\
e_{i, 13}\\
e_{i, 22}\\
\end{pmatrix}
=
\begin{pmatrix}
- \{a_{i-1}+b_{i}\}_{20} + 1 + \partial \ell(\theta) / \partial \theta_2 \\
-a_{i-1,21} \\
- b_{i,21} \\
-a_{i-1,22} - (e_{i,00} + 3) e^{-2\theta_2} \\
-b_{i,22} - (e_{i,00} + 3) e^{-2\theta_2}\\
2 \left(e_{i,00}+3\right)e^{-2\theta_2} \\
a_{i-1,21} e^{-2\theta_2}  \\
b_{i,21} e^{-2\theta_2} \\
\{- 2a_{i-1} + b_i\}_{21} e^{-2\theta_2} \\
\{-2 b_i + a_{i-1}\}_{21} e^{-2\theta_2} \\
(a_{i-1,22} + e^{-2\theta_2})e^{-2\theta_2}\\
(b_{i,22} + e^{-2\theta_2})e^{-2\theta_2}\\
-2 e^{-2\theta_2} (a_{i-1,22}  + 2e^{-2\theta_2})\\
-2 e^{-2\theta_2} (b_{i,22} + 2e^{-2\theta_2})\\
(\{a_{i-1}+b_i\}_{22} + 6 e^{-2\theta_2})e^{-2\theta_2} 
\end{pmatrix}
$$
and the exact value of each double integral involved in Equation~(\ref{eq:h22.2}) is provided with details in~\ref{app:I}.

\section{Results}\label{sec:results}

 In this section we illustrate the method over simulation studies using three different parameters $(\eta^\ast, \varepsilon^\ast) = (1.1, 0.2)$, $(\eta^\ast, \varepsilon^\ast) = (0.2,1.1)$ and $(\eta^\ast, \varepsilon^\ast) = (0.6, 0.2)$ and $\alpha$ (fixed) set to $\alpha^*=1$. A parameter used for simulation (respectively an estimate) is indicated with an asterisk (respectively a hat). All simulation studies are conducted using the R statistical programming environment \citep{rlanguage2024}.

%%%%%%%%%%%%%%%%%%%%%%%%
% RESULTS: GRADIENT AND OBSERVED FIM
%%%%%%%%%%%%%%%%%%%%%%%%

We firstly simulated $m=25$ random walks of length $n=100$ each within each simulation scheme and reported the score, ie. the gradient of the log-likelihood in Table~\ref{tab:grad} and the observed FIM, i.e. the negative of the Hessian of the log-likelihood in Table~\ref{tab:fim}. The score (respectively the observed information) is computed in its exact version using Equation~(\ref{eq:grad.ll}) (respectively Equations~(\ref{eq:h11}), (\ref{eq:h21}), (\ref{eq:h22.1}) and~(\ref{eq:h22.2})) and compared to an approximate using numerical differentiation of the log-likelihood recursively computed with Equation~(\ref{eq:fw}) (respectively numerical differentiation of the exact score) with a step size $h =10^{-6}$. Score and observed FIM are reported at points $\left(\log \eta^*, \log \varepsilon^* \right)$ and a random choice of $\left(\log 0.8, \log 0.6\right)$ and $\left(\log 0.1, \log 3\right)$. These last two points are arbitrary and other choices led to same conclusions. Additionally the observed information is also reported at parameter estimate $(\log \hat \eta, \log \hat \varepsilon)$ computed with the NR algorithm using exact values of the score and exact values of the observed FIM. The NR algorithm was initialized with the parameter associated to the best log-likelihood among a set of 10 initial parameters. As expected, the value of the score vanishes at $(\log \hat \eta, \log \hat \varepsilon)$ and therefore it is not reported at that point. We can see the concordance of the exact method when compared to the approximate one for both the score and the observed FIM at any point with the difference that the exact method provides exact results, is more stable and computationally more efficient than the approximate one.

\begin{table}
\setlength{\tabcolsep}{0.4em} 
\centering
\begin{tabular}{c | c | rr |rr | rr |}
\multicolumn{2}{l|}{\textbf{Simu. scheme} $\boldsymbol{\rightarrow}$}  & \multicolumn{2}{c}{$(\eta^*,\varepsilon^*)=(1.1,0.2)$}  & \multicolumn{2}{c}{$(\eta^*,\varepsilon^*) = (0.2,1.1)$} & \multicolumn{2}{c}{$(\eta^*,\varepsilon^*) = (0.6,0.2)$}\\[1ex] 
\cline{3-8}
\multicolumn{1}{c}{ \textbf{at} $\boldsymbol{\downarrow}$}  & 
\multicolumn{1}{c|}{\textbf{wrt} $\boldsymbol{\rightarrow}$}   & 
$\boldsymbol{\log \eta}$ & $\boldsymbol{\log \varepsilon}$ &  $\boldsymbol{\log \eta}$ & $\boldsymbol{\log \varepsilon}$ &  $\boldsymbol{\log \eta}$ & $\boldsymbol{\log \varepsilon}$ \\[0.8ex] 
\cline{2-8}
\addlinespace
\multirow{2}{*}{$\begin{pmatrix}\log\eta^* \\ \log\varepsilon^*\end{pmatrix}$} 
&\textbf{Exact} &  
 \bf 2.030 &\bf  5.775 & \bf - 2.341  &\bf  55.925 & \bf  9.701& \bf   9.137\\
&Approx.&    
 2.028  & 5.775 &   2.341 &  55.923 &   9.699  & 9.137\\[2ex] 
\multirow{2}{*}{$\begin{pmatrix}\log 0.8\\ \log 0.6\end{pmatrix}$} 
&\textbf{Exact} &   
 \bf 777.987  &\bf -103.242 &\bf   -224.430  &\bf  888.496 &\bf  -873.889 &\bf   -609.084\\
& Approx.&    
 777.985 &  -103.242 &   -224.431  & 888.494  &  -873.890 &  -609.085\\[2ex]   
\multirow{2}{*}{$\begin{pmatrix}\log 0.1\\ \log 3\end{pmatrix}$}
& \textbf{Exact} &
 \bf -3.213 &\bf  -1807.244 &\bf  -4.686 &  -2129.747 & \bf -4.811 &\bf   -2282.037\\
& Approx.& 
 -3.213 &  -1807.245 &   -4.686  & -2129.747&   -4.811&   -2282.037\\
\bottomrule
\end{tabular}
\caption{Score computed at different points $\theta = (\log \eta, \log \varepsilon)$ noted in the first header column (by pair of rows) over datasets composed of $m=25$ random walks of length $n=100$ each, simulated with parameter $(\eta^*,\varepsilon^*)$ indicated in the first header row. It is reported as partial derivative with respect to (wrt) $\log \eta$ and $\log \varepsilon$, per column (see second header row) and computed in its exact version using Equation~(\ref{eq:grad.ll}) and an approximate using numerical differentiation of the log-likelihood  with a step size $h = 10^{-6}$ (per row, see second header column).}
\label{tab:grad}
\end{table}

\begin{sidewaystable}
\centering
\setlength{\tabcolsep}{0.2em} 
\centering
    \begin{tabular}{l|l | rrr | rrr | rrr}
\multicolumn{2}{l|}{\textbf{Simu. scheme} $\boldsymbol{\rightarrow}$}  
& \multicolumn{3}{c}{$(\eta^*,\varepsilon^*)=(1.1,0.2)$}  
& \multicolumn{3}{c}{$(\eta^*,\varepsilon^*) = (0.2,1.1)$} 
& \multicolumn{3}{c}{$(\eta^*,\varepsilon^*) = (0.6,0.2)$} \\[1ex] 
\multicolumn{2}{l|}{\textbf{Estimate} $\boldsymbol{\rightarrow}$}  
& \multicolumn{3}{c}{$(\hat\eta,\hat\varepsilon) = (1.099, 0.205)$}  
& \multicolumn{3}{c}{$(\hat\eta,\hat\varepsilon) = (0.175, 1.122)$} 
& \multicolumn{3}{c}{$(\hat\eta,\hat\varepsilon) = (0.600, 0.204)$}\\[1ex] 
\cline{3-11}
\addlinespace
\multicolumn{1}{c}{ \textbf{at} $\boldsymbol{\downarrow}$} 
& \multicolumn{1}{c|}{\textbf{wrt} $\boldsymbol{\rightarrow}$}    
& $\begin{pmatrix}\log\eta\\\log\eta\end{pmatrix}$ & $\begin{pmatrix}\log\varepsilon\\\log\eta\end{pmatrix}$ & $\begin{pmatrix}\log\varepsilon\\\log\varepsilon\end{pmatrix}$ 
& $\begin{pmatrix}\log\eta\\\log\eta\end{pmatrix}$ & $\begin{pmatrix}\log\varepsilon\\\log\eta\end{pmatrix}$ & $\begin{pmatrix}\log\varepsilon\\\log\varepsilon\end{pmatrix}$ 
& $\begin{pmatrix}\log\eta\\\log\eta\end{pmatrix}$ & $\begin{pmatrix}\log\varepsilon\\\log\eta\end{pmatrix}$ & $\begin{pmatrix}\log\varepsilon\\\log\varepsilon\end{pmatrix}$ \\[2ex]
    \cmidrule{2-8}
    \multirow{2}{*}{$\begin{pmatrix}\log\hat\eta\\\log\hat\varepsilon\end{pmatrix}$} 
    & \textbf{Exact} & 
    \bf 4305.566 & \bf 227.885 & \bf 238.664  & \bf 15.08 & \bf  210.967 & \bf  4562.986& \bf 3766.322 & \bf 392.504  & \bf 448.671\\
    &Approx.& 
    4305.562  & 227.885 &  238.664  & 15.08 &  210.967 &  4562.982 & 3766.319 &  392.503  & 448.671\\[2ex]
    \multirow{2}{*}{$\begin{pmatrix}\log\eta^*\\\log\varepsilon^*\end{pmatrix}$} 
    &\textbf{Exact} &\bf 4321.495 & \bf  227.827 & \bf 238.46 & \bf  22.375  & \bf 281.925 &\bf   4530.306 &  \bf 3798.932 &  \bf 395.041 &  \bf 448.663\\
    &Approx.& 
 4321.492  & 227.827 &  238.46 &  22.375  & 281.926 &  4530.303 & 3798.929 &  395.040 &  448.663\\[2ex]
    \multirow{2}{*}{$\begin{pmatrix}\log 0.8\\\log 0.6\end{pmatrix}$} 
    & \textbf{Exact} & 
    \bf 3600.384 & \bf 1194.276 &\bf  360.553&  \bf 1692.59  & \bf 1097.237  & \bf 2441.067 &  \bf 1633.120 &\bf   -142.211  & \bf 685.356\\
    &Approx.& 
      3600.383  & 1194.274 &  360.553 &  1692.59  & 1097.236 &  2441.067& 1633.120 &  -142.212  & 685.356\\[2ex]
    \multirow{2}{*}{$\begin{pmatrix}\log 0.1\\\log 3\end{pmatrix}$} 
    &\textbf{Exact} &
    \bf 6.424 & \bf -1.832 & \bf 1376.325 & \bf  9.347 & \bf  -7.700 & \bf  737.188 & \bf  9.595  &\bf  -8.197 &\bf   433.104\\
    &Approx.&     
    6.424 &  -1.832  & 1376.324 &  9.346  & -7.700 &  737.187& 9.595  & -8.197 &  433.103\\
    \bottomrule
    \end{tabular}    
\caption{
Observed Fisher information computed at different points $\theta = (\log \eta, \log \varepsilon)$ noted in the first header column (by pair of rows) over datasets composed of $m=25$ random walks of length $n=100$ each, simulated with parameter $(\eta^*,\varepsilon^*)$ indicated in first header row.
The parameter estimate $(\hat\eta,\hat\varepsilon)$, computed with the Newton-Raphson algorithm using exact values of the score and the observed information is provided in the second header row.
The observed Fisher information is reported, per column, as partial derivatives with respect to $(\log \eta, \log \eta)$, $(\log \eta, \log \varepsilon)$ and $(\log \varepsilon, \log \varepsilon)$ (see second group of header rows), in its exact version using  Equations~(\ref{eq:h11}), (\ref{eq:h21}), (\ref{eq:h22.1}) and~(\ref{eq:h22.2}) and an approximate using numerical differentiation of the exact value of the score (see Equation~(\ref{eq:score})) with a step size $h = 10^{-6}$ (per row, see second header column).}
 \label{tab:fim}
\end{sidewaystable}

%%%%%%%%%%%%%%%%%%%%%
% RESULTS: PARAMETER ESTIMATES
%%%%%%%%%%%%%%%%%%%%%

We secondly report, in Table~\ref{tab:estimates}, an exemple of use of the observed FIM with parameter estimates computed with the NR algorithm using exact score and observed FIM and, additionally, their 95\% confidence intervals computed with the inverse of the exact observed FIM. Datasets were composed of different numbers $m$ of random walks ($m=25$, $m=100$ and $m=225$) of length $n=100$ each and simulated within each simulation scheme, i.e. $(\eta^*,\varepsilon^*) = (1.1, 0.2)$, $(\eta^*,\varepsilon^*) = (0.2,1.1)$ and $(\eta^*,\varepsilon^*) = (0.6, 0.2)$. 
We can notice as expected that confidence intervals shrink in expected proportions, that is around twice (respectively three times) for four times (respectively nine times) more sequences. 

\begin{table}
\centering
\begin{tabular}{l | r | ccc }
\toprule
$(\eta^*,\varepsilon^*) \downarrow $ & \multicolumn{1}{c}{} &$m=25$ &$m=100$ & $m=225$ \\
\midrule
 \multirow{2}{*}{$(1.1,0.2)$} & $\hat \eta$ & 
 1.099   [1.066;   1.133] &  1.111   [1.094;   1.128] &  1.116   [1.105;   1.128]\\
& $\hat \varepsilon$ &  
 0.205   [0.180;   0.234] &   0.193   [0.181;   0.207]&  0.196   [0.188;   0.205]\\
\addlinespace
 \multirow{2}{*}{$(0.2,1.1)$} & $\hat \eta$ & 
 0.175   [0.075;   0.409]&  0.235   [0.186;   0.297] & 0.205   [0.168;   0.251]\\
 & $\hat \varepsilon$   & 
 1.122   [1.068;   1.178]&  1.090   [1.063;   1.117] &  1.100   [1.082;   1.118]\\
 \addlinespace
  \multirow{2}{*}{$(0.6,0.2)$} & $\hat \eta$ &  
  0.600   [0.581;   0.621] &   0.607   [0.597;   0.617]&    0.609   [0.602;   0.616]\\
 & $\hat \varepsilon$   &  
 0.204   [0.185;   0.225] &    0.195   [0.185;   0.205] &  0.199   [0.192;   0.206]\\

 \bottomrule
\end{tabular}
\caption{Parameter estimates, computed with the NR algorithm using exact values of the score and the observed FIM, along with their 95\% confidence intervals over datasets simulated with different number $m$ of random walks ($m=25$, $m=100$ and $m=225$, in columns) and different ``true'' parameter $(\eta^*,\varepsilon^*)$ (per pair of rows). For each dataset, $\alpha=1$ and all chains are of length $n=100$.}
\label{tab:estimates}
\end{table}

%%%%%%%%%%%%%%%%%%%
% GRAPHICAL REPRESENTATION
%%%%%%%%%%%%%%%%%%%

\begin{figure}
\centering 
\begin{subfigure}{0.33\textwidth}
\includegraphics[width=\linewidth]{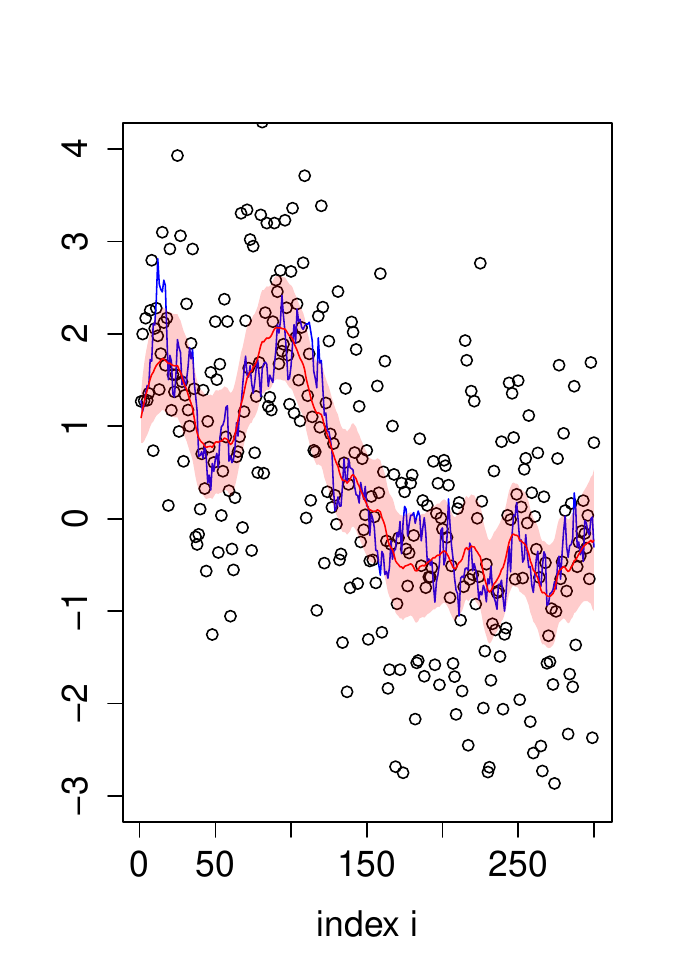}
\centering  \caption*{$(\eta^*,\varepsilon^*) = (1.1, 0.2)$ \\ $(\hat\eta,\hat\varepsilon) = (1.088, 0.149)$}
\end{subfigure}\hfil 
\begin{subfigure}{0.33\textwidth}
\includegraphics[width=\linewidth]{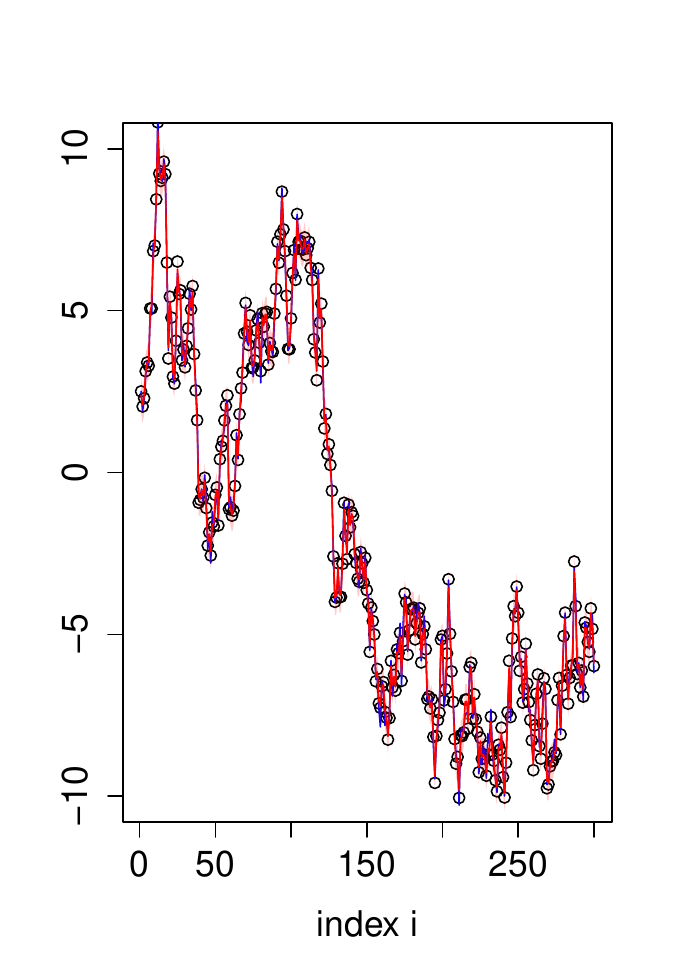}
\caption*{$(\eta^*,\varepsilon^*) = (0.2, 1.1)$ \\ $(\hat\eta,\hat\varepsilon) = (0.301, 1.044)$}
\end{subfigure}\hfil 
\begin{subfigure}{0.33\textwidth}
\includegraphics[width=\linewidth]{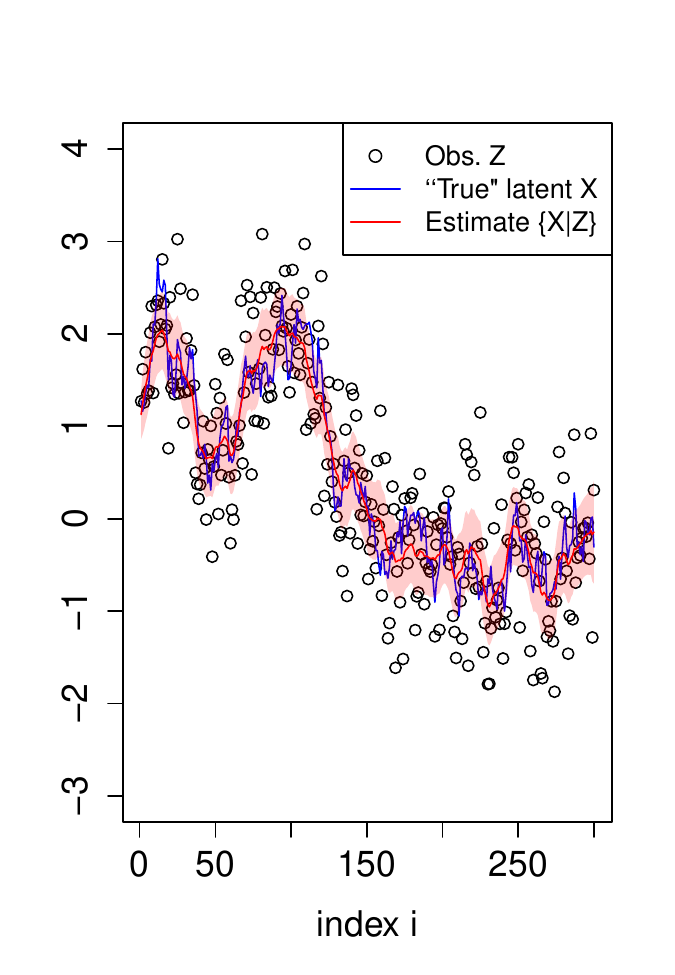}
\caption*{$(\eta^*,\varepsilon^*) = (0.6, 0.2)$ \\ $(\hat\eta,\hat\varepsilon) = (0.602, 0.153)$}
\end{subfigure}\hfil 
\caption{Graphical representation of a simulated random walk of length $n=300$ simulated with different ``true'' parameter $(\eta^*,\varepsilon^*)$ reported in sub-captions and $\alpha$ set to 1. Parameter estimates $(\hat\eta,\hat\epsilon)$, computed with the NR algorithm, are also reported in sub-captions. Note that the range of the y-axis differs between the middle panel and the other two. For each dataset, we represented the observed vector $Z = (z_1,\ldots,z_{300})$ (black dots), the ``true'' (simulated) latent random walk $X = (x_1,\ldots,x_{300})$ (blue line) and its estimated density as posterior mean $(M_1,\ldots,M_{300})$ (red line) and 95\% confidence interval computed with its estimated standard deviation $(S_1,\ldots,S_{300})$ (red polygon) computed according to Equation~(\ref{eq:post.x}). The similar shape is explained by the fact that simulations were seeded with the same R seed and conducted in the same order, starting with the entire latent random walk $X$, then the observed sequence $Z$.}
\label{fig:dataset}
\end{figure}

We finally propose, in Figure~\ref{fig:dataset}, a graphical representation of one sequence of length $n=300$ simulated with ``true'' parameter $(\eta^*,\varepsilon^*) = (1.1, 0.2)$, $(\eta^*,\varepsilon^*) = (0.2,1.1)$ and $(\eta^*,\varepsilon^*) = (0.6, 0.2)$ reported in sub-captions. 
For each simulation scheme, we represented the latent (simulated) random walk $X$, the observed sequence $Z$ and the estimated vector of posterior mean of $X$ along with its 95\% confidence interval computed with the estimated vector of standard deviation of $X$ according to Equation~(\ref{eq:post.x}). Parameter estimates were computed with the NR algorithm over the single sequence and initialized with the parameter associated to best log-likelihood among a set of ten random ones.
Standard deviations $(S_i)_{i=1,\ldots,300}$ display a plateau within each simulation scheme along the majority of the sequence except the first and the last one to ten indices with rising SD at the beginning and the end of the sequence.
The similar shape of all three random walks is explained by the fact that each simulation was seeded with the same R seed and conducted in the same order, firstly the entire latent random walk $X$, secondly the observed sequence $Z$. 
We can see that, as expected, the greater the parameter $\varepsilon^*$, the wider the range of the y-axis, the greater the parameter $\eta^*$, the wider the dispersion of $Z$ around $X$. The figure displays accurate posterior state estimates and we notice that the variance of posterior states increases with both $\eta$ and $\varepsilon$ with a larger impact of $\eta$ in concordance with Equation~(\ref{eq:post.x}).

%%%%%%%%%%%%%%%%%%%%%
%DISCUSSION AND PERSPECTIVE
%%%%%%%%%%%%%%%%%%%%%

\section{Discussion and perspectives}\label{sec:discussion}

We propose in the is work an analytical and a closed-form expression of the score and the observed FIM in a hidden random walk with Gaussian observation noise based on Oakes' identity. A recursive implementation leads to the same complexity in time as for computing the likelihood, that is, linear in the length of the sequence. We restricted our method to the univariate Gaussian case for simplicity and our first perspective is its extension to the multivariate case and the Gaussian linear state-space model. Moreover the mean  $\alpha$ of the 
first latent variable is assumed to be fixed and it would be interesting to investigate the introduction to a random intercept. Additionally, the method involves the gradient of posterior state densities as an intermediate quantity that it would be interesting to study for alternative questions, for instance, the sensitivity of posterior state estimates to perturbations. 

\section{Acknowledgments}
Funding: This work was supported by \textit{REWIND: ANR-22-PESN-0016}. 

\section{Competing interests}
The authors have no competing interest to declare. 

\section{Code availability}
The code related to all this work is available at the link \url{https://github.com/AlexandraLefe/FIM_noisy_GRW}.

\appendix
%%%%%%%%%%%%%%%%%%%%%%%%%%%%%%%%%
% PROOF OF FW AND BK RECURSION
%%%%%%%%%%%%%%%%%%%%%%%%%%%%%%%%%

\section{Proof of Equations~(\ref{eq:fw}) and~(\ref{eq:bk}): forward and backward quantities} \label{app:proof.eq.fw.bk}

Both equations are proved by induction, let us start with Equation~(\ref{eq:fw}). Substituting initial and emission densities in Equation~(\ref{eq:f1}) by those of the random walk written in Equation~(\ref{eq:local.densities}), the initial forward quantity becomes $f_1(x) = g(x,\alpha,\varepsilon) g(x,z_1,\eta)$ and therefore
$$
f_1(x) = 
g(z_1,\alpha,\sqrt{\varepsilon^2+\eta^2})
\, g\left(x,\frac{z_1\varepsilon^2+\alpha\eta^2}{\varepsilon^2+\eta^2},
\sqrt{\frac{\varepsilon^2\eta^2}{\varepsilon^2+\eta^2}}
\right)
$$
by a direct application of Equation~(\ref{eq:prod}). Hence we have $f_1(x) = e^{K_1} g(x,\mu_1,\sigma_1)$ with
$$
K_1=\log g(z_1,\alpha,\sqrt{\varepsilon^2+\eta^2}), \quad
\mu_1=\frac{z_1\varepsilon^2+\alpha\eta^2}{\varepsilon^2+\eta^2} 
\quad\text{and}\quad
\sigma_1=\sqrt{\frac{\varepsilon^2\eta^2}{\varepsilon^2+\eta^2}}.
$$
By induction, for $i=2,\ldots, n$, starting from Equation~(\ref{eq:fi}) and applying Equation~(\ref{eq:prod}) sequentially, we get
\begin{align*}
F_i(y)
& =\int e^{K_{i-1}} g(x,\mu_{i-1},\sigma_{i-1}) g(y,x,\varepsilon)g(z_i,y,\eta)dx \\
& =e^{K_{i-1}} g(z_i,y,\eta) \int g(x,\mu_{i-1},\sigma_{i-1}) g(x,y,\varepsilon)dx\\
& = e^{K_{i-1}} g(z_i,y,\eta) g(\mu_{i-1},y,\sqrt{\sigma_{i-1}^2+\varepsilon^2})\\
F_i(y)
& =e^{K_{i-1}} g(z_i,\mu_{i-1},\sqrt{\sigma_{i-1}^2+\varepsilon^2+\eta^2})
\, g\left(y,\frac{z_i(\sigma_{i-1}^2+\varepsilon^2)+(\mu_{i-1})\eta^2}{\sigma_{i-1}^2+\varepsilon^2+\eta^2},
\sqrt{\frac{(\sigma_{i-1}^2+\varepsilon^2)\eta^2}{\sigma_{i-1}^2+\varepsilon^2+\eta^2}}\right).
\end{align*}
Consequently we have $f_i(x) = e^{K_i} g(x,\mu_i,\sigma_i)$ with
$$
K_i = K_{i-1}+\log g(z_i,\mu_{i-1},\sqrt{\sigma_{i-1}^2+\varepsilon^2+\eta^2}),
$$
$$
\mu_i=\frac{z_i(\sigma_{i-1}^2+\varepsilon^2)+\mu_{i-1}\eta^2}{\sigma_{i-1}^2+\varepsilon^2+\eta^2}
\qquad \text{and} \qquad
\sigma_i=\sqrt{\frac{(\sigma_{i-1}^2+\varepsilon^2)\eta^2}{\sigma_{i-1}^2+\varepsilon^2+\eta^2}}
$$
which concludes the proof of Equation~(\ref{eq:fw}). 

\bigskip
Conducting a similar reasoning for the backward pass, we return to the initial backward quantity of Equation~(\ref{eq:bn}) and we can write $h_n(x) = e^{L_n}g(x,\nu_n,\tau_n)$ where
$$
L_n=-\log g(0,\nu_n,\tau_n)
\quad \text{with}\quad 
\nu_n=0
\quad \text{and} \quad 
\tau_n=+\infty.
$$ 
In practice, initializing the standard deviation $\tau_n$ at $\tau_n=10^5$ in sufficient in most cases for approximating $B_n(y)\simeq 1$ for all $y \in \mathbb R$. By induction for $i = n, \ldots, 2$ and applying Equation~(\ref{eq:prod}) sequentially,
\begin{align*}
h_{i-1}(x) 
& = e^{L_{i}} \int g(y,x,\varepsilon)g(y,z_i,\eta) g(y,\nu_{i},\tau_{i}) dy \\
& = e^{L_{i}} g(z_i,\nu_i,\sqrt{\eta^2+\tau_i^2})\int g(y,x,\varepsilon)
\, g \left(
y,\frac{\nu_i\eta^2+z_i\tau_i^2}{\eta^2+\tau_i^2},\sqrt{\frac{\eta^2\tau_i^2}{\eta^2+\tau_i^2}}
\right)dy\\
h_{i-1}(x)  & = e^{L_{i}}g(z_i,\nu_i,\sqrt{\eta^2+\tau_i^2}) 
\, g\left(x,
  \frac{\nu_i\eta^2+z_i\tau_i^2}{\eta^2+\tau_i^2}-\delta,
  \sqrt{\varepsilon^2 + \frac{\eta^2\tau_i^2}{\eta^2+\tau_i^2}}
\right)
.
\end{align*}
Therefore we can recursively compute 
\begin{equation*}
L_{i-1} = L_i + \log g\left(z_i,\nu_i,\sqrt{\eta^2+\tau_i^2}\right), 
\quad
\nu_{i-1 }= \frac{\nu_i\eta^2+z_i\tau_i^2}{\eta^2+\tau_i^2} 
\quad \text{and} \quad 
\tau_{i-1}=\sqrt{\varepsilon^2 + \frac{\eta^2\tau_i^2}{\eta^2+\tau_i^2}}
\end{equation*}
which concludes the proof of Equation~(\ref{eq:bk}).

%%%%%%%%%%%%%%%%%%%%%%%%%%%%%%%%%
% PROOF OF POST X AND POST XY
%%%%%%%%%%%%%%%%%%%%%%%%%%%%%%%%%

\section{Proof of Equations~(\ref{eq:post.x}) and~(\ref{eq:post.xy}): posterior state densities}\label{app:proof.eq.post.xy}
Replacing FB quantities in the joint density of $\{X_i, Z=z\}$ as expressed in Equation~(\ref{eq:joint.x}) by their analytical expressions in Equations~(\ref{eq:fw}) and~(\ref{eq:bk}), we can write $\mathcal P(X_i=x, Z=z | \theta) = e^{K_i+L_i} g(x,\mu_i,\sigma_i)g(x,\nu_i,\tau_i)$. We simply apply Equation~(\ref{eq:prod}) to obtain 
$$
\mathcal P(X_i=x, Z=z | \theta)
= e^{K_i+L_i} \, g\left (\mu_i, \nu_i, \sqrt{\sigma_i^2 + \tau_i^2}\right) g\left(x, M_i,S_i\right)
$$
with $M_i$ and $S_i$ detailed in Equation~(\ref{eq:post.x}). 
Dividing this expression by its integration over $x$, that is $e^{K_i+L_i} \, g\left (\mu_i, \nu_i, \sqrt{\sigma_i^2 + \tau_i^2}\right)$, we get $\mathcal P(X_i=x|Z=z; \theta)$ and we conclude the proof of Equation~(\ref{eq:post.x}). 

\bigskip
Likewise, the proof of Equation~(\ref{eq:post.xy}) starts from Equation~(\ref{eq:joint.xy}) in which FB quantities are replaced by their expressions in  Equations~(\ref{eq:fw}) and~(\ref{eq:bk}) such that the joint density of $\{X_{i-1}, X_i, Z=z\}$ writes
$$
\mathcal{P}(X_{i-1}=x, X_i=y, Z=z|\theta) = e^{K_{i-1}+L_i} g(x,\mu_{i-1},\sigma_{i-1}) g(y,x,\varepsilon)g(z_i,y,\eta) g(y,\nu_i,\tau_i).
$$
Applying Equation~(\ref{eq:prod}) sequentially, this equation becomes 
\begin{align*}
\mathcal{P}(X_{i-1}=x, &X_i=y, Z=z| \theta) \\
 &= e^{K_{i-1}+L_i} 
g\left(y, \mu_{i-1}, \sqrt{\varepsilon_i^2+\sigma_{i-1}^2}\right)
g\left(x, \frac{\mu_{i-1} \varepsilon^2+y\sigma_{i-1}^2}{\varepsilon^2+\sigma_{i-1}^2}, \sqrt{\frac{\sigma_{i-1}^2\varepsilon^2}{\sigma_{i-1}^2+\varepsilon^2} }\right)\\
& \hspace{11em}
\times
g\left(\nu_i, z_i, \sqrt{\tau_i^2+\eta^2}\right)
g\left(y, \frac{\nu_i\eta^2+z_i\tau_i^2}{\eta^2+\tau_i^2}, \sqrt{\frac{\eta^2\tau_i^2}{\eta^2+\tau_i^2}}\right) \\
& = e^{K_{i-1}+L_i} 
g\left(\nu_i, z_i, \sqrt{\tau_i^2+\eta^2}\right)
g\left(\frac{\nu_i\eta^2+z_i\tau_i^2}{\eta^2+\tau_i^2}, \mu_{i-1},  \sqrt{\frac{\eta^2\tau_i^2}{\eta^2+\tau_i^2} + \varepsilon_i^2+\sigma_{i-1}^2}\right) \\
& \hspace{20em}
\times
g\left(x, P_iy+Q_i, R_i \right) g\left(y,N_i, T_i\right)
\end{align*}
where $P_i$, $Q_i$, $R_i$, $N_i$ and $T_i$ are detailed in Equation~(\ref{eq:post.xy}). 
Dividing this last equation by the likelihood that is its the double integration over $x$ and $y$, we conclude the proof of Equation~(\ref{eq:post.xy}). 

%%%%%%%%%%%%%%%%%%%%%%%%%%%%%%%%%
% PROOF OF FB GRADIENTS
%%%%%%%%%%%%%%%%%%%%%%%%%%%%%%%%%

\section{Proof of Equation~(\ref{eq:grad.fb}): Gradient of FB quantities}\label{app:proof.eq.grad}

Equations~(\ref{eq:grad.fb}) is proved, in this appendix, by induction. Let us firstly highlight the fact that for any $\mu,\sigma\in\mathbb R$, the partial derivative of $\log g(x,\mu,\sigma)$ with respect to $\log \sigma$ is given by 
\begin{equation}\label{eq:partial.g}
\partial\log g(x,\mu,\sigma) / \partial \log \sigma = -1+(x-\mu)^2e^{-2\log\sigma}
\end{equation}
where $g(\cdot,\mu,\sigma)$ is the Gaussian density of mean $\mu$ and standard deviation $\sigma$.
We also remind that the parameter of interest is $\theta = (\theta_1, \theta_2) = (\log \eta, \log \varepsilon)$.

\paragraph{Gradient of forward quantities}
Differentiating Equation~(\ref{eq:f1}) in which densities $\rho$ and $\varphi_1$ (respectively their gradient) are replaced by those of the Gaussian random walk of Equation~(\ref{eq:local.densities}) (respectively their analytical expression provided in Equation~(\ref{eq:partial.g})), the gradient of the initial forward quantity writes
$$
\nabla f_1(x) =  
\begin{pmatrix}
-1+(x-z_1)^2e^{-2\theta_1}\\
-1+(x-\alpha)^2e^{-2\theta_2}
\end{pmatrix}
f_1(x) 
= A_{1} \left(x^k\right)_{k=0}^2 \ f_1(x)
$$
where
$$
A_1 = 
\begin{pmatrix}
-1 + z_1^2e^{-2\theta_1} & - 2 z_1 e^{-2\theta_1} & e^{-2\theta_1} \\
-1 + \alpha^2e^{-2\theta_2} & - 2 \alpha e^{-2\theta_2} & e^{-2\theta_2}
\end{pmatrix}.
$$
Moreover, differentiating Equation~(\ref{eq:fi}) in which densities are replaced by those of the Gaussian random walk, the gradient of the $i^\text{th}$ forward quantity writes 
\begin{align*}
\nabla f_i(y) 
& =
\int 
\left\{\nabla f_{i-1}(x)\right\} g(y,x,\varepsilon) g(z_i,y,\eta) 
+ f_{i-1}(x) \left\{\nabla g(y,x,\varepsilon) g(z_i,y,\eta)\right\} \, dx \\
& =
\int 
\bigg[
\nabla f_{i-1}(x)
+ f_{i-1}(x) \{\nabla \log g(y,x,\varepsilon) + \nabla\log g(z_i,y,\eta)\}
\bigg]
g(y,x,\varepsilon) g(z_i,y,\eta) 
\ dx.
\end{align*}
Therefore, by induction, for $i = 2, \ldots, n$, we have
$$
\nabla f_i(y) = 
e^{K_{i-1}} 
\int 
\left[
A_{i-1} (x^k)_{k=0}^2 + 
\begin{pmatrix}
-1+(y-z_i)^2e^{-2\theta_1}\\
-1+(y-x)^2e^{-2\theta_2}
\end{pmatrix}
\right]
g(x,\mu_{i-1},\sigma_{i-1}) g(y,x,\varepsilon) g(z_i,y,\eta)  \,dx.
$$
after injecting Equations~(\ref{eq:fw}) and~(\ref{eq:partial.g}). 
Applying Equation~(\ref{eq:prod}) over the product of $g(x,\mu_{i-1},\sigma_{i-1})$ and $g(y,x,\varepsilon)$, we obtain 
\begin{multline}\label{eq:interm.grad.fw}
\nabla f_i(y) =
e^{K_{i-1}} g\left(z_i,y,\eta\right) g\left(\mu_{i-1},y,\sqrt{\sigma_{i-1}^2+\varepsilon^2}\right) \\
\times
\int \left[ A_{i-1} \left(x^k\right)_{k=0}^2 + 
\begin{pmatrix}
-1+(y-z_i)^2e^{-2\theta_1}\\
-1+(y-x)^2e^{-2\theta_2}
\end{pmatrix}
\right]
g\left(x,P_iy+Q_i, R_i\right) \, dx
\end{multline}
where $P_i$, $Q_i$ and $R_i$ are given in Equation~(\ref{eq:post.xy}). Note that the term before the integral of this equation (the term on the first line on the right hand side) is precisely $F_i(y)$. Indeed, one can start from Equation~(\ref{eq:fi}) in which densities are substituted by those of Equation~(\ref{eq:local.densities}), replace $f_{i-1}(x)$ by its expression in Equation~(\ref{eq:fw}), apply Equation~(\ref{eq:prod}) over the product $g(x,\mu_{i-1},\sigma_{i-1}) g(y,x,\varepsilon)$ and marginalize $x$ out to obtain $f_i(y)$. 

Gathering coefficients according to the powers of $x$ in Equation~(\ref{eq:interm.grad.fw}) and extracting the zero, first and the second order moments of the Gaussian distribution of mean $(P_iy+Q_i)$ and variance $R_i^2$, we can write
\begin{equation*}
\nabla f_i(y) =
f_i(y) \left[
A_{i-1} + 
\begin{pmatrix}
-1+(y-z_i)^2e^{-2\theta_1} & 0 & 0 \\
-1 + y^2 e^{-2\theta_2} & -2y e^{-2\theta_2} & e^{-2\theta_2}
\end{pmatrix}
\right]
\begin{bmatrix}
1 \\
P_iy+Q_i \\
(P_iy+Q_i)^2+R_i^2 
\end{bmatrix}.
\end{equation*}
Finally, gathering coefficients according to the powers of $y$, we conclude that 
\begin{equation*}
\nabla f_i(y) =
A_i \left(y^k\right)_{k=0}^2 \ f_i(y)
\end{equation*}
with 
\begin{multline*}
A_i = 
A_{i-1}
\begin{pmatrix}
1&0&0 \\
Q_i&P_i&0\\
Q_i^2+R_i^2& 2P_iQ_i& P_i^2
\end{pmatrix}\\
+
\begin{pmatrix}
-1 + z_i^2e^{-2\theta_1} & -2z_ie^{-2\theta_1} & e^{-2\theta_1}\\
-1 + (Q_i^2+R_i^2)e^{-2\theta_2} & 2(P_i-1)Q_ie^{-2\theta_2} & (P_i-1)^2e^{-2\theta_2}
\end{pmatrix}
\end{multline*}
which concludes the proof of the left part of Equation~(\ref{eq:grad.fb}) along with the recursive formulas for its implementation. 

\paragraph{Gradient of backward quantities}

With a similar reasoning but slightly more complicated formulas, following the recursive implementation of backward quantities reminded in Equations~(\ref{eq:bn}) and~(\ref{eq:bi}), the gradient of the initial backward can be expressed as 
$$
\nabla h_n(x) = B_n h_n(x)
$$
where $B_n$ is a zero matrix of size $2\times3$. 
Therefore, by induction, we have
\begin{align*}
\nabla h_{i-1}(x)   
& =
\int 
\left\{ \nabla h_{i}(y) \right\} g(y,x,\varepsilon) g(z_i,y,\eta) + h_{i}(y) \left\{\nabla g(y,x,\varepsilon) g(z_i,y,\eta) \right\}
\, dy
\\ 
& =  
e^{L_i}\int 
\left[ B_i (y^k)_{k=0}^2 + 
\begin{pmatrix}
-1+(y-z_i^2)e^{-2\theta_1} \\
-1+(y-x^2)e^{-2\theta_2}
\end{pmatrix}
\right] 
g(y,\nu_{i},\tau_{i}) g(y,x,\varepsilon) g(z_i,y,\eta) \ dy \\
\end{align*}
after injecting Equations~(\ref{eq:local.densities}) and~(\ref{eq:bk}) in Equations~(\ref{eq:bi}) and differentiating with respect to $\theta$.
Applying Equation~(\ref{eq:prod}) over the product $g(y,\nu_{i},\tau_{i})g(z_i,y,\eta)$ we can write
\begin{multline*}
\nabla h_{i-1}(x) 
= e^{L_{i}}g\left(z_i,\nu_i,\sqrt{\eta^2+\tau_i^2}\right)\\
\times
\int
\left[ 
B_i (y^k)_{k=0}^2 +
\begin{pmatrix}
-1+(y-z_i^2)e^{-2\theta_1} \\
-1+(y-x^2)e^{-2\theta_2}
\end{pmatrix}
\right] 
g(y,x,\varepsilon) g(y,\nu_{i-1},\omega_i) \,dy 
\end{multline*}
where $\nu_{i-1}$ and $\omega_i$ are detailed in Equations~(\ref{eq:bk}). Applying Equation~(\ref{eq:prod}) once more we get 
\begin{multline*}
\nabla h_{i-1}(x)   
= e^{L_{i}}g\left(z_i,\nu_i,\sqrt{\eta^2+\tau_i^2}\right)g\left(x, \nu_{i-1}, \sqrt{\varepsilon^2 + \omega_i^2}\right) \\ 
\times \int
\left[ B_i (y^k)_{k=0}^2 + 
\begin{pmatrix}
-1+(y-z_i^2)e^{-2\theta_1} \\
-1+(y-x^2)e^{-2\theta_2}
\end{pmatrix}
\right] 
g\left(y, \widetilde P_i x + \widetilde Q_i, \widetilde R_i\right) \,dy
\end{multline*}
with
$$
\widetilde P_i = \frac{\omega_i^2}{\varepsilon^2 + \omega_i^2}
\qquad
\widetilde Q_i = \frac{\nu_{i-1}\varepsilon^2}{\varepsilon^2 + \omega_i^2}
\qquad
\widetilde R_i=\sqrt{\widetilde P_i  \varepsilon^2}.
$$
We recognize $h_{i-1}(x)$ in the term before the integral when applying sequentially Equation~(\ref{eq:prod}) to $B_{i-1}(x) = e^{L_i} \int g(y,\nu_i,\tau_i)g(y,x,\varepsilon)g(z_i,y,\eta) \, dy$. Therefore we can write
\begin{align*}
\nabla h_{i-1}(x)   
& =
h_{i-1}(x)
\int
\left[ B_i (y^k)_{k=0}^2 + 
\begin{pmatrix}
-1+(y-z_i^2)e^{-2\theta_1} \\
-1+(y-x^2)e^{-2\theta_2}
\end{pmatrix}
\right] 
g\left(y, \widetilde P_i x + \widetilde Q_i, \widetilde R_i\right) \,dy.
\end{align*}
Gathering coefficients according to the powers of $y$ and extracting the zero, first and second order moment of the Gaussian distribution of mean $\widetilde P_i x + \widetilde Q_i$ and variance $\widetilde R_i^2$, the equation becomes
\begin{align*}
\nabla h_{i-1}(x)   
& =
h_{i-1}(x) 
\left[
B_i + 
\begin{pmatrix}
-1+z_i^2 e^{-2\theta_1} & -2 z_i e^{-2\theta_1} & e^{-2\theta_1} \\
-1+x^2 e^{-2\theta_2} & -2x e^{-2\theta_2} & e^{-2\theta_2}
\end{pmatrix}
\right]
\begin{bmatrix}
1 \\
\widetilde P_ix+\widetilde Q_i \\
(\widetilde P_ix+\widetilde Q_i)^2+\widetilde R_i^2 
\end{bmatrix}.
\end{align*}
Finally, gathering coefficients according to the powers of $x$, we conclude that
$$
\nabla h_{i-1}(x) = B_{i-1} \left(x^k\right)_{k=0}^2 \ h_{i-1}(x)
$$
with 
\begin{multline*}
B_{i-1} = 
B_i
\begin{pmatrix}
1&0&0 \\
\widetilde Q_i&\widetilde P_i&0\\
\widetilde Q_i^2+\widetilde R_i^2& 2\widetilde P_i\widetilde Q_i& \widetilde P_i^2
\end{pmatrix} \\
+
\begin{pmatrix}
-1 + \left[(\widetilde Q_i-z_i)^2+\widetilde R_i^2\right]e^{-2\theta_1} & 2\widetilde P_i(\widetilde Q_i-z_i)e^{-2\theta_1} & \widetilde P_i^2e^{-2\theta_1}\\
-1 + (\widetilde Q_i^2+\widetilde R_i^2)e^{-2\theta_2} & 2(\widetilde P_i-1)\widetilde Q_ie^{-2\theta_2} & (\widetilde P_i-1)^2e^{-2\theta_2}
\end{pmatrix}
\end{multline*}
which concludes the right part of Equation~(\ref{eq:grad.fb}). 

%%%%%%%%%%%%%%%%%%%%%%%%%%%%%%%%%
% PROOF OF GRADIENT OF POST X AND POST XY
%%%%%%%%%%%%%%%%%%%%%%%%%%%%%%%%%

\section{Proof of Equations~(\ref{eq:grad.post.x}) and~(\ref{eq:grad.post.xy}): Gradient of posterior state densities}\label{app:proof.grad.post}

The proof of Equation~(\ref{eq:grad.post.x}) is a simple combination of Equations~(\ref{eq:joint.x}), (\ref{eq:grad.post}) and~(\ref{eq:grad.fb}).  
Replacing $\bar X$ by $X_i$ in Equation~(\ref{eq:grad.post}), we have
\begin{equation}\label{eq:interm.proof.partial.postx}
\nabla \mathcal P(X_i=x | Z=z; \theta) = 
\frac{\nabla \mathcal P(X_i=x, Z=z| \theta)}{L(\theta)} 
- \mathbb{P}(X_i=x | Z=z; \theta) \nabla\ell(\theta).
\end{equation}
As reminded in Equation~(\ref{eq:joint.x}), we have $\mathcal P(X_i=x,Z=z|\theta) = f_i(x)h_i(x)$. Therefore its gradient, that is the numerator of the first part of the right hand side of Equation~(\ref{eq:interm.proof.partial.postx}), writes
$
\nabla \mathcal P(X_i=x, Z=z| \theta) = 
\nabla f_i(x) h_i(x) + f_i(x) \nabla h_i(x).
$
Replacing the gradient of FB quantities by their expression in Equations~(\ref{eq:grad.fb}), we obtain
$$
\nabla \mathcal P(X_i=x, Z=z| \theta) =
f_i(x) h_i(x) \left(A_{i} + B_{i}\right) \left(x^k\right)_{k=0}^2.
$$
We recognize the joint probability $\mathcal P(X_i=x, Z=z | \theta)$ in the element before the braces of the right hand side of this equation, that is $f_i(x) h_i(x)$. Therefore, Equation~(\ref{eq:interm.proof.partial.postx}) becomes 
\begin{equation*}
\nabla \mathcal P(X_i=x | Z=z; \theta) =
 \mathcal P(X_i=x | Z=z; \theta) \left[ (A_{i} + B_{i}) (x^k)_{k=0}^2 - \nabla\ell(\theta)\right]
\end{equation*}
We finally replace the posterior density $\mathcal P(X_i=x | Z=z; \theta)$ by its expression in Equation~(\ref{eq:post.x}) to conclude the proof of Equation~(\ref{eq:grad.post.x}). 

\bigskip
The proof of Equation~(\ref{eq:grad.post.xy}) is very similar. Starting with Equation~(\ref{eq:grad.post}) in which $\bar X$ is replaced by $\{X_{i-1}, X_i\}$, we can write 
\begin{multline}\label{eq:interm.proof.partial.postxy}
\nabla \mathcal P(X_{i-1}=x, X_i=y | Z=z; \theta) =
\frac{
\nabla \mathcal P(X_{i-1}=x, X_i=y, Z=z| \theta)
}{
L(\theta)} \\
- \mathcal P(X_{i-1}=x, X_i=y | Z=z; \theta)
\nabla \ell(\theta).
\end{multline}
Based on the analytical expression of the joint probability of $\{X_{i-1}=x, X_i=y, Z=z\}$ reminded in Equation~(\ref{eq:joint.xy}), its gradient, that is the numerator of the first part of the right hand side of Equation~(\ref{eq:interm.proof.partial.postxy}), writes 

\begin{align*}
\nabla \mathcal P(&X_{i-1}=x, X_i=y, Z=z| \theta) \\ 
&=  
\nabla f_{i-1}(x) g(y,x,\varepsilon) g(y,z_i,\eta) h_i(y) \\
&\hspace{6em}
+ f_{i-1}(x)
\begin{bmatrix}
\partial\log g(y,z_i,\eta)/\partial \theta_1\\
\partial\log g(y,x,\varepsilon)/\partial \theta_2
\end{bmatrix}
 g(y,x,\varepsilon) g(y,z_i,\eta) h_i(y)   \\
& \hspace{14em} 
+ f_{i-1}(x) g(y,z_i,\eta)  g(y,x,\varepsilon)\nabla h_i(y) \\ 
& = 
f_{i-1}(x) g(y,x,\varepsilon) g(y,z_i,\eta) h_i(y)
 \left(A_{i-1} (x^k)_{k=0}^2 
 +
 \begin{bmatrix}
- 1+(y-z_i)^2 e^{-2\theta_1} \\
- 1+(y-x)^2 e^{-2\theta_2}
\end{bmatrix} 
+ B_i (y^k)_{k=0}^2  \right)\\
& = 
\mathcal P(X_{i-1} = x, X_i=y, Z=z|\theta) 
 \left(A_{i-1} (x^k)_{k=0}^2 
 +
 \begin{bmatrix}
- 1+(y-z_i)^2 e^{-2\theta_1} \\
- 1+(y-x)^2 e^{-2\theta_2}
\end{bmatrix} 
+ B_i (y^k)_{k=0}^2  \right)
.
\end{align*}
We inject this expression in Equation~(\ref{eq:interm.proof.partial.postxy}) and we replace the posterior density $\mathcal P(X_{i-1}=x,  X_i=y | Z=z; \theta)$ by its expression in Equation~(\ref{eq:post.xy}) to conclude the proof of Equation~(\ref{eq:grad.post.xy}). 

\section{Closed-form solutions for the integrals involved in Equation~(\ref{eq:h22.2})}\label{app:I}
We provide, in this appendix, closed-form solutions for the integrals involved in Equation~(\ref{eq:h22.2}) where index $i$ is dropped from the notation to alleviate it. We denote by $\mathrm I_{jk}$ the double integral $\mathrm I_{jk} = \int \int x^jy^k g(x,Py + Q , R) g(y,N,T) \,dxdy$ for any constant $P, Q, R, N, T$. 
Starting with $\mathrm I_{40}$ we have 
\begin{align*}
\mathrm I_{40} & = \int \int x^4 g(x, Py+Q, R) g(y,N,T) \, dxdy \\
& = \int \left[ (Py+Q)^4 + 6 (Py+Q)^2R^2+3R^4  \right] g(y,N,T) \, dy\\
\mathrm I_{40} & = P^4 \mathcal M_4 + 4P^3Q \mathcal M_3 + 6P^2\left(Q^2+R^2\right) \mathcal M_2 + (4PQ^3+12PQR^2) \mathcal M_1+ Q^4+6R^2Q^2+3R^4
\end{align*}
where $\mathcal M_{4} = N^4+6N^2T^2+3T^4$, $\mathcal M_{3} = N^3+3NT^2$, $ \mathcal M_{2} = N^2 + T^2$ and $ \mathcal M_{1} = N$ are respectively the fourth, third, second and first moments of the Gaussian distribution of mean $N$ and standard deviation $T$. 
With a similar reasoning we get 
\begin{align*}
\mathrm I_{30} & = P^3 \mathcal M_3 + 3P^2Q \mathcal M_2 + 3P (Q^2+R^2) \mathcal M_1+ Q^3+3QR^2 \\
\mathrm I_{20} & = P^2 \mathcal M_2 + 2PQ \mathcal M_1+ Q^2+R^2 \\
\mathrm I_{10} & = P \mathcal M_1+ Q
\end{align*}

Pursuing with $\mathrm I_{31}$ we can write 
\begin{align*}
\mathrm I_{31} & = \int \int x^3y \, g(x, Py+Q, R) g(y,N,T) \, dxdy \\
& = \int y \left[ (Py+Q)^3 + 3 (Py+Q)R^2 \right] g(y,N,T) \, dy\\
& = \int y \left[ P^3y^3 + 3P^2Qy^2 + 3PQ^2y + Q^3 + 3PR^2y + 3QR^2 \right] g(y,N,T) \, dy\\
\mathrm I_{31} & = P^3 \mathcal M_4 + 3P^2Q \mathcal M_3 + 3P\left(Q^2+R^2\right) \mathcal M_2 + (Q^3+3QR^2) \mathcal M_1
\end{align*}
and similarly 
\begin{align*}
\mathrm I_{22} & = P^2  \mathcal M_4 + 2PQ \mathcal M_3 + (Q^2+R^2) \mathcal M_2 \\
\mathrm I_{21} & = P^2  \mathcal M_3 + 2PQ \mathcal M_2 + (Q^2+R^2) \mathcal M_1 \\
\mathrm I_{13} & = P  \mathcal M_4 + Q\mathcal M_3 \\
\mathrm I_{12} & = P  \mathcal M_3 + Q\mathcal M_2 \\
\mathrm I_{11} & = P  \mathcal M_2 + Q\mathcal M_1.
\end{align*}

Finally we have $\mathrm I_{04} = \int\int y^4 \, g(x, Py+Q, R) g(y,N,T) \, dxdy = \mathcal M_4$
and likewise $\mathrm I_{03} = \mathcal M_3$, $\mathrm I_{02} = \mathcal M_2$, $\mathrm I_{01} = \mathcal M_1$ and $\mathrm I_{00}=1$.

%%%%%%%%%%%%%%%%%%%%%%%%%%%%%%%%%
% BIBLIOGRAPHY
%%%%%%%%%%%%%%%%%%%%%%%%%%%%%%%%%

  \bibliographystyle{elsarticle-num-names} 
  \bibliography{library.bib}

\begin{thebibliography}{17}
\expandafter\ifx\csname natexlab\endcsname\relax\def\natexlab#1{#1}\fi
\providecommand{\url}[1]{\texttt{#1}}
\providecommand{\href}[2]{#2}
\providecommand{\path}[1]{#1}
\providecommand{\DOIprefix}{doi:}
\providecommand{\ArXivprefix}{arXiv:}
\providecommand{\URLprefix}{URL: }
\providecommand{\Pubmedprefix}{pmid:}
\providecommand{\doi}[1]{\href{http://dx.doi.org/#1}{\path{#1}}}
\providecommand{\Pubmed}[1]{\href{pmid:#1}{\path{#1}}}
\providecommand{\bibinfo}[2]{#2}
\ifx\xfnm\relax \def\xfnm[#1]{\unskip,\space#1}\fi
%Type = Article
\bibitem[{Baum and Petrie(1966)}]{baum1966statistical}
\bibinfo{author}{L.~E. Baum}, \bibinfo{author}{T.~Petrie},
\newblock \bibinfo{title}{{Statistical Inference for Probabilistic Functions of
  Finite State Markov Chains}},
\newblock \bibinfo{journal}{The Annals of Mathematical Statistics}
  \bibinfo{volume}{37} (\bibinfo{year}{1966}) \bibinfo{pages}{1554--1563}.
  \URLprefix \url{http://www.jstor.org/stable/2238772}.
  \DOIprefix\doi{https://doi.org/10.1214/aoms/1177699147}.
%Type = Article
\bibitem[{Baum et~al.(1970)Baum, Petrie, Soules, and
  Weiss}]{baum1970maximization}
\bibinfo{author}{L.~E. Baum}, \bibinfo{author}{T.~Petrie},
  \bibinfo{author}{G.~Soules}, \bibinfo{author}{N.~Weiss},
\newblock \bibinfo{title}{A maximization technique occurring in the statistical
  analysis of probabilistic functions of markov chains},
\newblock \bibinfo{journal}{The Annals of Mathematical Statistics}
  \bibinfo{volume}{41} (\bibinfo{year}{1970}) \bibinfo{pages}{164--171}.
  \URLprefix \url{http://www.jstor.org/stable/2239727}.
  \DOIprefix\doi{https://doi.org/10.1214/aoms/1177697196}.
%Type = Article
\bibitem[{Dempster et~al.(1977)Dempster, Laird, and
  Rubin}]{dempster1977maximum}
\bibinfo{author}{A.~P. Dempster}, \bibinfo{author}{N.~M. Laird},
  \bibinfo{author}{D.~B. Rubin},
\newblock \bibinfo{title}{{Maximum Likelihood from Incomplete Data via the EM
  Algorithm}},
\newblock \bibinfo{journal}{Journal of the Royal Statistical Society. Series B
  (Methodological)} \bibinfo{volume}{39} (\bibinfo{year}{1977})
  \bibinfo{pages}{1--38}. \URLprefix \url{http://www.jstor.org/stable/2984875}.
  \DOIprefix\doi{https://doi.org/10.1111/j.2517-6161.1977.tb01600.x}.
%Type = Inproceedings
\bibitem[{Orchard and Woodbury(1972)}]{orchard1972missing}
\bibinfo{author}{T.~Orchard}, \bibinfo{author}{M.~A. Woodbury},
\newblock \bibinfo{title}{{A missing information principle: theory and
  applications}},
\newblock in: \bibinfo{booktitle}{Proceedings of the Sixth Berkeley Symposium
  on Mathematical Statistics and Probability, Volume 1: Theory of Statistics},
  volume~\bibinfo{volume}{6}, \bibinfo{organization}{University of California
  Press}, \bibinfo{year}{1972}, pp. \bibinfo{pages}{697--716}.
  \DOIprefix\doi{https://doi.org/10.1525/9780520325883-036}.
%Type = Article
\bibitem[{Louis(1982)}]{louis1982finding}
\bibinfo{author}{T.~A. Louis},
\newblock \bibinfo{title}{{Finding the Observed Information Matrix when Using
  the EM Algorithm}},
\newblock \bibinfo{journal}{Journal of the Royal Statistical Society. Series B
  (Methodological)} \bibinfo{volume}{44} (\bibinfo{year}{1982})
  \bibinfo{pages}{226--233}. \URLprefix
  \url{http://www.jstor.org/stable/2345828}.
  \DOIprefix\doi{https://doi.org/10.1111/j.2517-6161.1982.tb01203.x}.
%Type = Article
\bibitem[{Oakes(1999)}]{oakes1999direct}
\bibinfo{author}{D.~Oakes},
\newblock \bibinfo{title}{Direct calculation of the information matrix via the
  em algorithm},
\newblock \bibinfo{journal}{Journal of the Royal Statistical Society. Series B
  (Statistical Methodology)} \bibinfo{volume}{61} (\bibinfo{year}{1999})
  \bibinfo{pages}{479--482}. \URLprefix
  \url{http://www.jstor.org/stable/2680653}.
  \DOIprefix\doi{https://doi.org/10.1111/1467-9868.00188}.
%Type = Article
\bibitem[{Turner et~al.(1998)Turner, Cameron, and Thomson}]{turner1998hidden}
\bibinfo{author}{T.~R. Turner}, \bibinfo{author}{M.~A. Cameron},
  \bibinfo{author}{P.~J. Thomson},
\newblock \bibinfo{title}{Hidden markov chains in generalized linear models},
\newblock \bibinfo{journal}{The Canadian Journal of Statistics / La Revue
  Canadienne de Statistique} \bibinfo{volume}{26} (\bibinfo{year}{1998})
  \bibinfo{pages}{107--125}. \URLprefix
  \url{http://www.jstor.org/stable/3315677}.
  \DOIprefix\doi{https://doi.org/10.2307/3315677}.
%Type = Article
\bibitem[{Delyon et~al.(1999)Delyon, Lavielle, and
  Moulines}]{delyon1999expectation}
\bibinfo{author}{B.~Delyon}, \bibinfo{author}{M.~Lavielle},
  \bibinfo{author}{E.~Moulines},
\newblock \bibinfo{title}{Convergence of a stochastic approximation version of
  the em algorithm},
\newblock \bibinfo{journal}{The Annals of Statistics} \bibinfo{volume}{27}
  (\bibinfo{year}{1999}) \bibinfo{pages}{94--128}. \URLprefix
  \url{http://www.jstor.org/stable/120120}.
  \DOIprefix\doi{https://doi.org/10.1214/aos/1018031103}.
%Type = Article
\bibitem[{Chalmers(2018)}]{chalmers2018numerical}
\bibinfo{author}{R.~P. Chalmers},
\newblock \bibinfo{title}{{Numerical approximation of the observed information
  matrix with Oakes' identity}},
\newblock \bibinfo{journal}{British Journal of Mathematical and Statistical
  Psychology} \bibinfo{volume}{71} (\bibinfo{year}{2018})
  \bibinfo{pages}{415--436}. \URLprefix
  \url{https://bpspsychub.onlinelibrary.wiley.com/doi/abs/10.1111/bmsp.12127}.
  \DOIprefix\doi{https://doi.org/10.1111/bmsp.12127}.
%Type = Inproceedings
\bibitem[{Capp{\'e} and Moulines(2005)}]{cappe2005recursive}
\bibinfo{author}{O.~Capp{\'e}}, \bibinfo{author}{E.~Moulines},
\newblock \bibinfo{title}{{Recursive computation of the score and observed
  information matrix in hidden Markov models}},
\newblock in: \bibinfo{booktitle}{IEEE/SP 13th Workshop on Statistical Signal
  Processing, 2005}, \bibinfo{organization}{IEEE}, \bibinfo{year}{2005}, pp.
  \bibinfo{pages}{703--708}. \DOIprefix\doi{10.1109/SSP.2005.1628685}.
%Type = Article
\bibitem[{Bartolucci and Farcomeni(2014)}]{bartolucci2015information}
\bibinfo{author}{F.~Bartolucci}, \bibinfo{author}{A.~Farcomeni},
\newblock \bibinfo{title}{{Information matrix for hidden Markov models with
  covariates}},
\newblock \bibinfo{journal}{Statistics and Computing} \bibinfo{volume}{25}
  (\bibinfo{year}{2014}) \bibinfo{pages}{515--526}. \URLprefix
  \url{https://link.springer.com/article/10.1007/s11222-014-9450-8}.
  \DOIprefix\doi{10.1007/s11222-014-9450-8}.
%Type = Article
\bibitem[{Lystig and Hughes(2002)}]{lystig2002exact}
\bibinfo{author}{T.~C. Lystig}, \bibinfo{author}{J.~P. Hughes},
\newblock \bibinfo{title}{Exact computation of the observed information matrix
  for hidden markov models},
\newblock \bibinfo{journal}{Journal of Computational and Graphical Statistics}
  \bibinfo{volume}{11} (\bibinfo{year}{2002}) \bibinfo{pages}{678--689}.
  \URLprefix \url{http://www.jstor.org/stable/1391119}.
  \DOIprefix\doi{10.1198/106186002402}.
%Type = Inproceedings
\bibitem[{Lefebvre and Nuel(2018)}]{lefebvre2018sum}
\bibinfo{author}{A.~Lefebvre}, \bibinfo{author}{G.~Nuel},
\newblock \bibinfo{title}{{A sum-product algorithm with polynomials for
  computing exact derivatives of the likelihood in Bayesian networks}},
\newblock in: \bibinfo{editor}{V.~Kratochvíl}, \bibinfo{editor}{M.~Studený}
  (Eds.), \bibinfo{booktitle}{Proceedings of the Ninth International Conference
  on Probabilistic Graphical Models}, volume~\bibinfo{volume}{72} of
  \textit{\bibinfo{series}{Proceedings of Machine Learning Research}},
  \bibinfo{publisher}{PMLR}, \bibinfo{year}{2018}, pp.
  \bibinfo{pages}{201--212}. \URLprefix
  \url{https://proceedings.mlr.press/v72/lefebvre18a.html}.
%Type = Article
\bibitem[{Rabiner(1989)}]{rabiner1989tutorial}
\bibinfo{author}{L.~Rabiner},
\newblock \bibinfo{title}{{A tutorial on hidden Markov models and selected
  applications in speech recognition}},
\newblock \bibinfo{journal}{Proceedings of the IEEE} \bibinfo{volume}{77}
  (\bibinfo{year}{1989}) \bibinfo{pages}{257--286}.
  \DOIprefix\doi{10.1109/5.18626}.
%Type = Book
\bibitem[{Capp{\'e} et~al.(2005)Capp{\'e}, Moulines, and
  Ryd{\'e}n}]{cappe2005inference}
\bibinfo{author}{O.~Capp{\'e}}, \bibinfo{author}{E.~Moulines},
  \bibinfo{author}{T.~Ryd{\'e}n}, \bibinfo{title}{{Inference in hidden Markov
  models}}, \bibinfo{publisher}{Springer}, \bibinfo{year}{2005}.
  \DOIprefix\doi{10.1007/0-387-28982-8}.
%Type = Techreport
\bibitem[{Bromiley(2003)}]{bromiley2003products}
\bibinfo{author}{P.~A. Bromiley}, \bibinfo{title}{Products and Convolutions of
  Gaussian Probability Density Functions}, \bibinfo{type}{Internal Report}
  \bibinfo{number}{TINA Memo No. 2003-003}, University of Manchester,
  \bibinfo{year}{2003}.
%Type = Manual
\bibitem[{{R Core Team}(2024)}]{rlanguage2024}
\bibinfo{author}{{R Core Team}}, \bibinfo{title}{{R: A Language and Environment
  for Statistical Computing}}, \bibinfo{organization}{R Foundation for
  Statistical Computing}, \bibinfo{address}{Vienna, Austria},
  \bibinfo{year}{2024}. \URLprefix \url{https://www.R-project.org/}.

\end{thebibliography}
\end{document}